\documentclass[twocolumn,showpacs,preprintnumbers,amsmath,amssymb]{revtex4}
\usepackage{graphicx}% Include figure files
\usepackage{dcolumn}% Align table columns on decimal point
\usepackage{bm}% bold math
\begin{document}
\preprint{APS/123-QED}

\title{Nematic textures in spherical shells.}% Force line breaks with \\

\author{V. Vitelli and D. R. Nelson}%
\affiliation{%
Department of Physics, Harvard University, Cambridge MA, 02138
}%

\date{\today}

\begin{abstract}
The equilibrium texture of nematic shells is studied as a function of their thickness. 
For ultrathin shells the ground
state has four short $\frac{1}{2}$ disclination lines but, as the
thickness of the film increases, a three dimensional escaped
configuration composed of two pairs of half-hedgehogs becomes
energetically favorable. We derive an exact solution for the
nematic ground state in the one Frank constant approximation and
study the stability of the corresponding texture against thermal
fluctuations. 
\end{abstract}

\pacs{Valid PACS appear here}% PACS, the Physics and Astronomy
\maketitle

\section{\label{sec:intr}Introduction}

The study of liquid crystal phases benefits from geometrical
reasoning in two important ways. Firstly, liquid crystal
elasticity can often be cast in terms of the curvature of
equipotential lines (or surfaces) that map out the corresponding
textures. Second, the observed textures are strongly affected by
geometric and topological constraints imposed by the presence of
boundaries confining the system. The liquid crystal ground state
results from the competition between the energetic requirement of
minimizing the "curvature of the texture" and the geometric
frustration introduced by boundaries that impart a preferred
curvature at the edge of the sample that often cannot propagate
across the system \cite{deGennesbook,Kleman-book,MKleman}.

The boundary conditions can be controlled experimentally with the
possibility of designing molecular systems with intriguing
technological applications \cite{Kami03}. For example, colloidal
particles coated by a very thin nematic layer have in their ground
state four disclinations sitting at the vertices of a tetrahedron. 
Each coated colloidal particle can then
be viewed as the fundamental building block of a self assembled
lattice with tetravalent coordination. The "bonds" between the
colloidal particles could be provided by chemical linkers attached
at the four "bald spots" at the cores of the four disclinations
present in each colloid \cite{Nels02}. A second example, is provided
by self-assembled systems of block copolymers \cite{sega03} which
are a promising tool for "soft lithography" on both flat and curved substrates \cite{Park-Chaikin97}.
In addition, liquid crystals in confined geometries provide an
arena for physicists and mathematicians interested in 
applications of geometrical and topological ideas to material
science \cite{Kami02,Lavr98,Craw96}.

In this work we present a theoretical study of liquid crystal
phases (focusing on vector, nematic and hexatic order) confined in a spherical shell of
varying thickness with the director assumed to be tangent to the
two interfaces. We first consider the two dimensional regime
where a nematic film coats a quenched 
spherical surface such as a colloidal particle in
solution or the interface of, say, a water droplet in oil. The presence of topological defects in the ground
state for ordered states on spherical surfaces is unavoidable \cite{Nels83,Mack91,lube92} . 
Recent experimental and theoretical investigations of
spherical crystallography have provided an alternative context to
study the constraints posed by the compactness of the underlying
curved space \cite{bowi2000,baus03}.  More recent explorations have
concentrated on 2D ordered phases confined to interfaces of varying Gaussian curvature
\cite{Bowi03,ViteNels04,ViteTurn05} as well as dynamically fluctuating
surfaces \cite{Park96,Lenz03}.

\begin{figure}
\includegraphics[width=0.3\textwidth]{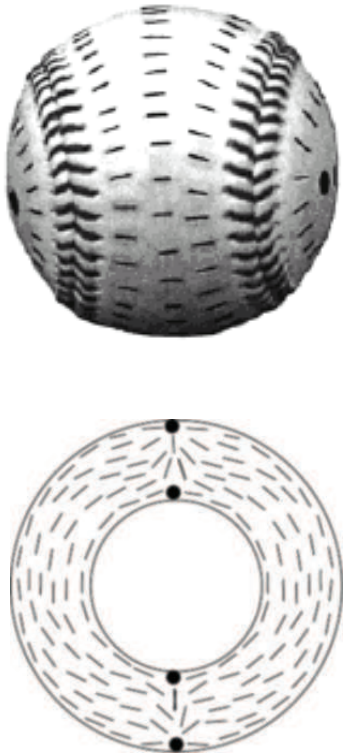}
\caption{\label{fig:summary}(Top panel) Two-dimensional texture characterized by four short disclination lines at the vertices of a tetrahedron inscribed in the sphere. The surface texture shown (inscribed on 
the surface of a baseball) is invariant throughout the thickness of the shell.
(Bottom panel) Cut view of the escaped three dimensional texture given by two pairs of half hedgehogs located at the north and south poles of the sphere.}
\end{figure}

As the thickness of a nematic film increases, an escaped three dimensional texture,
also strongly influenced by the spherical topology and the boundary conditions,
become energetically favored with respect to planar textures. This
instability destabilizes the tetravalent nematic texture on colloids. In this paper, we estimate 
the thickness of the nematic film above which a
texture with four radial disclination lines of charge $s=\frac{1}{2}$ becomes unstable to four half-hedgehogs. The two competing textures studied in this paper are shown in Fig. \ref{fig:summary}. We also discuss the possibility of hysteresis between the two textures.
%%%%%%%%%%%%%%%%%%%%%%%%%%%%%%%
%%%%%%%%%%%%%%%%%%%%%%%%%%%%%%%

The organization of this paper is as follows. In section
\ref{sec:texture} we derive exact solutions for the ground state
of spherical films of tilted molecules and nematogens 
%and hexatic phase %
within isotropic
elasticity by using the method of conformal mappings. 
In the notation of  References \cite{Nels02,lube92}, these situations correspond to
order parameters described by a bond angle with $p=1,2$ fold symmetry in the tangent plane of the sphere (see Appendix \ref{appA}). 
A mathematical justification for our approach is provided in
Appendix \ref{appB}, where the same technique is illustrated in the
context of a more familiar flat space problem. In section
\ref{sec:stability} we study the stability of liquid crystal
textures to thermal fluctuations by means of a normal mode
analysis whose details are relegated to Appendices \ref{appA} and
\ref{appC}. The stability of the valence-four texture against
escaped solutions is considered in section
\ref{sec:valence} where a phase diagram is derived with the
thickness as a control parameter. The
texture distortions caused by the elastic anisotropy between bend
and splay deformations are briefly considered in section \ref{sec:conc}.

\section{\label{sec:texture}Textures}

The liquid crystal free energy for molecules embedded in an
arbitrary frozen surface can be written in the one constant
approximation as
\begin{equation}
F=\frac{K}{2}\int dA \!\!\!\!\! \quad D_{i}n^{j}({\bf
u})D^{i}n_{j}({\bf u}) \!\!\!\! \quad , \label{free energy}
\end{equation}
where ${\bf u}=\{u_{1},u_{2}\}$ is a set of internal coordinates,
${\bf n}({\bf u})$ is the liquid crystal director defined in the
tangent plane, $D_{i}$ is the covariant derivative with respect to
the metric of the surface and $dA$ is the infinitesimal surface
area \cite{Nels87,Davi87,Park96,Davidreview}. The free energy 
of Eq.(\ref{free energy}) is invariant upon rotating each molecule 
\textbf{n}$(u)$ by the same (arbitrary) 
angle with respect to any axis of rotation perpendicular to the local
tangent plane. The treatment of
systems with a p-fold symmetry is straightforward provided that
the one Frank constant approximation is used for $p=1$ and $p=2$ and the consequences
of any additional couplings to curvature neglected \cite{Davi87}. This choice of
free energy implies that the minimal energy configuration will be
given locally by neighboring ${\bf n}({\bf u})$ vectors which
differ only by parallel transport. The curvature of the surface
induces "frustration" in the texture. In fact, by Gauss' "Theorema
egregium" \cite{Kami02}, tangent vectors parallel transported
along a closed loop are rotated by an amount equal to the Gaussian
curvature integrated over the enclosed area. On a sphere, this theorem 
insures that the nematic ground state always has four excess
disclinations \cite{Nels83,lube92}. More generally, the sum of the topological
charges on any closed surface is equal to the integrated Gaussian
curvature, implying a minimum of $2$ and $6$ disclinations in the ground state of
tilted molecules and hexatics, respectively. 
 
We introduce a local angle field $\alpha({\bf u})$,
corresponding to the angle between ${\bf n}({\bf u})$ and an
arbitrary local reference frame, we can rewrite the free
energy introduced in Eq.(\ref{free energy}) as:
\begin{equation}
F = \frac{1}{2}K\int dS \!\!\!\!\! \quad g^{ij}
(\partial_{i}\alpha - A_{i})(\partial_{j}\alpha - A_{j})\!\!\!\!
\quad , \label{eq:patic-ener}
\end{equation}
where $dS=d^{2}u\sqrt{g}$, $g$ is the determinant of the metric
tensor $g_{ij}$ and $A_{i}$ is the spin-connection whose curl is
the Gaussian curvature $G({\bf u})$ \cite{Davidreview,Kami02}. 
On a sphere of radius $R$ parametrized by polar coordinates $(\theta,\phi)$, 
the only non vanishing components 
of the (inverse) metric tensor are $g^{rr}=\frac{1}{R \sin \theta}$ and 
$g^{\phi \phi}=\frac{1}{R}$. A convenient choice of of the spin connection 
(which plays the role of the vector potential) is discussed in Appendix \ref{appA}.
The simplified free energy in Eq.(\ref{eq:patic-ener}) is the starting
point of our analysis.

\subsection{\label{tilt}Tilted molecules on a sphere}

The orientational order of molecules tilted by a constant angle
with respect to a spherical interface can be modelled by a vector
field $\textbf{n}(\theta,\phi)$ defined in the local tangent plane 
on which the molecule has a fixed length projection
\cite{Mack91}. To determine the ground state of the liquid crystal
texture, we minimize the Frank free energy of
Eq.(\ref{eq:patic-ener}). As discussed above, 
the topological charges must sum up to $4\pi$, the integrated
Gaussian curvature of the sphere \cite{Davidreview,Kami02}. For a vector field ($p=1$) the texture
with only two defects of charges $+2\pi$ minimizes the Frank free
energy and satisfies the topological constraint. Since the defects
repel each other they preferentially sit at two antipodal points
that we can designate as the north and south pole of the sphere.
If the splay and bend coupling constants of the nematic are equal, then there is a
large degeneracy in the ground state arising from the invariance
of the vector free energy in Eq.(\ref{eq:patic-ener}) under global
rotations $\alpha(\bf{u})$$\rightarrow$$\alpha(\bf{u})$+c, where $\textbf{u}\equiv{\theta,\phi}$.
One representative texture is a "sink" and a
"source" of $\textbf{n}(\textbf{u})$ at the two poles. In this splay rich texture $\textbf{n}(\textbf{u})$
is parallel to the lines of longitude on a sphere. In a bend rich texture, related to the previous by a $\frac{\pi}{2}$ rotation about the local normal to the surface, $\textbf{n}(\textbf{u})$ is everywhere parallel to the lines of latitude.
Any other rotation of $\textbf{n}(\textbf{u})$ that makes an arbitrary constant angle with respect to this texture is an acceptable solution for the ground state of the molecules. 

As we now show, this
degeneracy is lifted when $K_{3}\neq K_{1}$.
Indeed the effect of distinct splay and bend elastic
constants $K_1$ and  $K_3$ (the twist elastic constant $K_{2}$ is absent in two dimensions) is to select the bend-rich texture if $K_1>K_{3}$ or the splay-rich one if $K_3>K_{1}$. The intermediate
configurations obtained by a global rotation of the director are
now unstable. Assume for simplicity that $K_3>K_{1}$. In this
case, it is convenient to recast the Frank free energy (see Appendix A) as follows:
\begin{eqnarray}
F=\frac{1}{2} \int d^2 \textbf{x} \!\!\!\! \quad \sqrt{g} \!\!\!\!
\quad [K_1 \left(D_i \!\!\!\!\! \quad n^j\left) \!\!\!\! \quad
\right( D^i \!\!\!\!\! \quad n^j \right) \nonumber \\ + (K_3- K_1)
(\textbf{D} \times \textbf{n})^2 ] \!\!\!\! \quad , \label{eq:frank1}
\end{eqnarray}
where the covariant derivatives is expressed in terms of the
Christoffel connection, $\Gamma \!\!\!\!\! \quad ^{j}_{i
\!\!\!\!\!\!\! \quad t}$ ,
\begin{equation}
D_i \!\!\!\!\! \quad n^j = \partial _{i} \!\!\!\!\! \quad n^j +
\Gamma \!\!\!\!\! \quad  ^{j}_{i \!\!\!\!\!\!\! \quad t} \!\!\!\!
\quad n^t \!\!\!\!\! \quad , \label{eq:frank1b}
\end{equation}
and the covariant form of the curl squared is \cite{Davidreview,Weinberg-GRbook},
\begin{equation}
(\nabla \times \textbf{n})^2 \equiv \left(D_i \!\!\!\!\! \quad n_j
- D_j \!\!\!\!\! \quad n_i \left) \!\!\!\! \quad \right( D^i
\!\!\!\!\! \quad n^j - D^j \!\!\!\!\! \quad n^i \right)  \!\!\!\!
\quad . \label{eq:frank2}
\end{equation}
The first term in Eq.(\ref{eq:frank1}) resembles the Frank free
energy in the one coupling constant approximation and is minimized
by choosing the sink-source (or "lines of longitude") solution. The second term (which is
positive definite) will vanish for this texture since the sink-source texture is
bend free. All other textures have a higher energy. 

A similar argument can be used to prove that the two
vortex-configuration which follows the lines of latitude is the minimum of the free energy when
$K_1>K_{3}$ by rewriting the Frank free energy as
\begin{eqnarray}
F=\frac{1}{2} \int d^2 \textbf{x} \!\!\!\! \quad \sqrt{g} \!\!\!\!
\quad [K_3 \left(D_i \!\!\!\!\! \quad n^j\left) \!\!\!\! \quad
\right( D^i
\!\!\!\!\! \quad n_j \right) \nonumber\\
+ \left(K_1- K_3 \right) (\textbf{D} \cdot \textbf{n})^2 ] \!\!\!\!
\quad , \label{eq:frank3}
\end{eqnarray}
where the covariant form of the divergence reads
\begin{equation}
\textbf{D} \cdot \textbf{n} \equiv \frac{1}{\sqrt{g}}\partial _i
\left(\sqrt{g} \!\!\!\!\! \quad n^i \right)  \!\!\!\! \quad .
\label{eq:frank2}
\end{equation}
The latitudinal texture minimizes the first term of Eq.(\ref{eq:frank3}) while the second
vanishes because this texture is splay free. Any deviation from the splay-free latitudinal 
texture will only increase the energy. 

\begin{figure}
\includegraphics[width=0.45\textwidth]{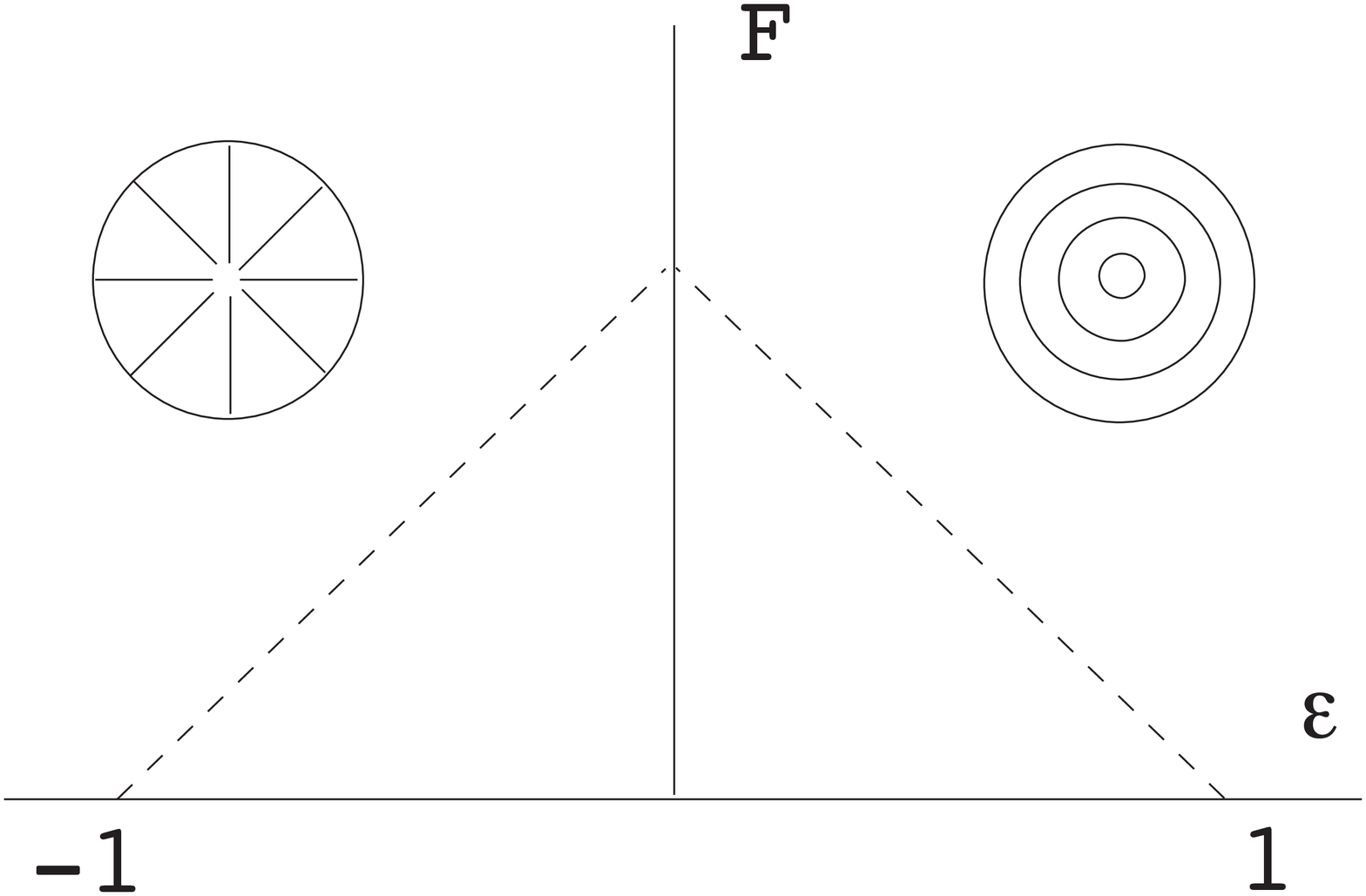}
\caption{\label{fig:tilted}Schematic illustration of the phase
diagram of the texture of tilted molecules on a sphere as a
function of the anisotropy parameter $\epsilon$ superimposed on a
plot of the free energy, $F$, stored in the texture versus
$\epsilon$. The two competing ground states (top view) are
characterized by either pure bend (lines of longitude configuration at left) or 
pure splay (lines of latitude configuration at right).}
\end{figure}

The energy of both textures can be expressed as a function of the
anisotropy parameter, $\epsilon$, and the mean of the elastic
constants, $K$,
\begin{eqnarray}
\epsilon &\equiv& \frac{K_3 - K_1}{K_3 + K_1}  \!\!\!\! \quad , \\
K &\equiv& \frac{K_3 + K_1}{2}  \!\!\!\! \quad ,
\label{eq:epsilon}
\end{eqnarray}
and the radius of the sphere, $R$, scaled by the short distance cutoff $a$. 
The resulting free energy for arbitrary $\epsilon$ reads (see Eq.(\ref{eq:frankexplicit3}))
\begin{equation}
F = 2 \pi K \left(1-|\epsilon \!\!\!\! \quad
|\right)\left(\ln\left[\frac{R}{a}\right]-0.3\right) \!\!\!\!
\quad . \label{eq:frank2}
\end{equation}
The conclusions of this section are summarized in Fig.
\ref{fig:tilted} which suggests that there is a discontinuous
first order transition when $\epsilon$ passes through zero. This analysis
mirrors similar arguments valid in the plane \cite{chandra-book}.
%%%%%%%%%%%%%%%%%%%%%%%%%%%%%%%
%%%%%%%%%%%%%%%%%%%%%%%%%%%%%%%

\subsection{\label{nem} Nematic Texture}

The nematic texture of very thin spherical shells of nematic liquid crystal with tangential boundary conditions can be analyzed
within the one Frank constant approximation by using the method of
conformal mappings whose mathematical justification is illustrated
in Appendix \ref{appB} by means of a simpler example.

An elegant argument introduced by Lubensky and Prost in
Ref.\cite{lube92} shows that the ground state of nematogens on a
sphere is given by $4$ disclinations of topological charge $s=1/2$ sitting at the
vertexes of a tetrahedron. The energy of
single disclinations is proportional to the square of its
strength. As a result, the longitudinal and latitudinal textures derived for tilted molecules in
section \ref{tilt} are unstable since their energies can be lowered by
splitting each $s=1$ defect at the north and south pole into two $s=1/2$
disclinations and letting them relax to their equilibrium positions at
the vertexes of a tetrahedron where they are as far away from each
other as possible. According to a calculation in Ref.\cite{lube92}, the
energy $F_s$ of a sphere of radius $R$ with in plane orientational
order and $2n$ interacting minimal disclinations for a p-fold order parameter
is given by 
\begin{eqnarray}
F_s = 2 \pi K \!\!\!\! \quad h \left[\frac{1}{p}\ln
\left(\frac{4 p^2 R}{a}\right) + c_p \right] \!\!\!\! \quad .
\label{eq:e4bis}
\end{eqnarray}
where the $\{c_{p}\}$ are constants depending on the symmetry of the order parameter and the defect core energy while $h$ is the thickness of the liquid crystal layer. The numerical values of the relevant constants are $c_1=0$
and $c_2\simeq - 0.2$. When $p=2$ is chosen in Eq.(\ref{eq:e4bis})
the elastic energy is indeed smaller than the corresponding value
for $p=1$ in the limit $R\gg a$, in agreement with related arguments given in Ref.\cite{Nels02}.

\begin{figure}
\includegraphics[width=0.3\textwidth]{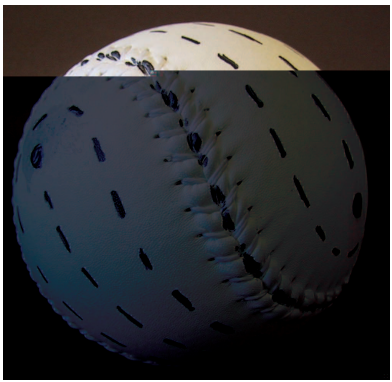}
\caption{\label{fig:baseball}Schematic illustration of the baseball texture of a thin nematic shell. The same texture is reproduced from a different perspective in the top panel of Fig. \ref{fig:summary}.}
\end{figure}

To obtain an algebraic expression for the texture we proceed as
illustrated in Appendix \ref{appB} and seek a function 
$\Omega(x,y,z)= \Phi(x,y,z)+ i \Psi(x,y,z)$ which is harmonic on
the sphere except for two arcs connecting the defects in pairs. The calculation 
for nematogens described below was suggested to us by F. Dyson \cite{Dyson}. The function
$\Phi(x,y,z)$, which we can interpret as an electrostatic
potential, takes equal and opposite values on the two arcs and is
equal to zero on a baseball-like seam (see Fig. \ref{fig:baseball}) which divides the sphere
into two congruent regions. 
%%%%%%%%%%%%%%%%%%%%%%%%%%%%%%%
%%%%%%%%%%%%%%%%%%%%%%%%%%%%%%%
The nematic director is then oriented
(up to a global rotation) along the contour lines of
$\Phi(x,y,z)$, that is the equipotential lines of this "curved
space capacitor". In this analogy, the contour lines of
$\Psi(x,y,z)$ are electric field lines, hence they correspond to a
valid texture where the director is rotated locally by $\frac{\pi}{2}$
with respect to the equipotential lines.  The arcs can be either
great-circle arcs extending more than half way round the sphere or
short great-circles arcs connecting the same pair of defects along
the shortest path. The first choice leads to equipotential lines
whose seam resembles in shape that of a baseball. If the second
choice is made the pattern of equipotential lines would not
deviate much from concentric circles and the seam would look more
like the seam of a cricket ball. We will explicitly show that the
two choices are equivalent since the equipotential lines of the
first solution are field lines of the second and vice versa.

\begin{figure}
\includegraphics[width=0.45\textwidth]{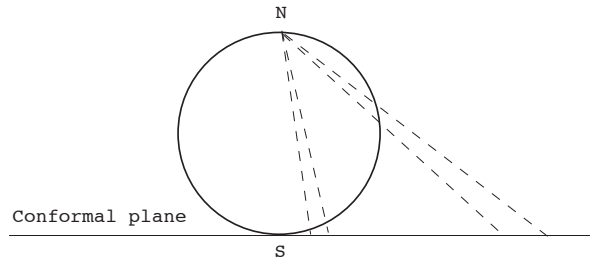}
\caption{\label{fig:stereo}Graphical construction of the
stereographic projection. Regions close to the north pole have
larger images in the conformal plane than regions of equal areas
close to the south pole. The stereographic projection preserves
the topology of the surface provided all points at infinity are
identified with the north pole.}
\end{figure}

\begin{figure}
\includegraphics[width=0.45\textwidth]{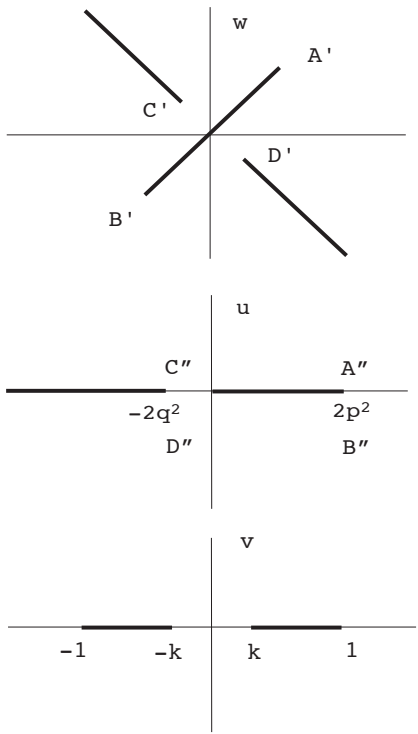}
\caption{\label{fig:mapping1}Illustration of the change in the
branch cut structures of the complex function $\Omega(v)$
describing the nematic texture after performing a series of
conformal transformations. In the top panel we have simply
performed a stereographic projection from the sphere. The middle panel shows the
"fold up" transformation of Eq.(\ref{eq:map1}) whereas the bottom
panel corresponds to the transformation in Eq.(\ref{eq:bilinear1})
that symmetrizes the positions of the cuts.}
\end{figure}

\begin{figure}
\includegraphics[width=0.45\textwidth]{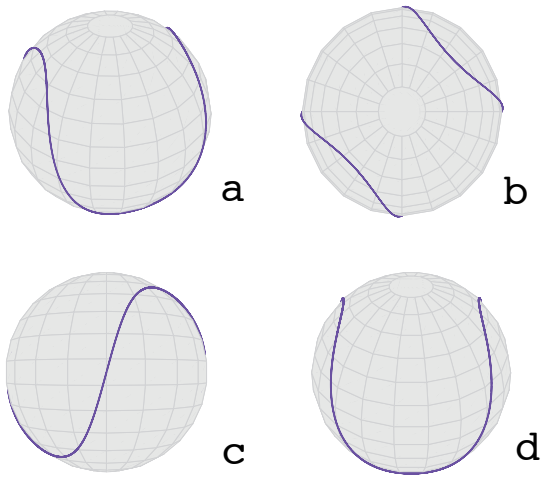}
\caption{\label{fig:seam3d} Different views of a track of parallel nematogens which partitions the sphere
into two equal areas, each containing two $s=\frac{1}{2}$ disclination defects. It
resembles a "fattened" version of the seam of a baseball.}
\end{figure}

We choose the arcs connecting the defect pairs along great circles and we
take the four defects labelled by A, B, C, D to lie at the
vertices of a tetrahedron inscribed on a sphere of radius $1$ and
whose north and south poles are $N=(0,0,1)$ and $S=(0,0,-1)$
respectively
\begin{eqnarray}
A&=&\frac{1}{\sqrt{3}}(1,1,1)   \!\!\! \quad , \!\! \quad
B=\frac{1}{\sqrt{3}}(-1,-1,1)   \!\!\! \quad , \nonumber\\
C&=&\frac{1}{\sqrt{3}}(-1,1,-1) \!\!\! \quad , \!\! \quad
D=\frac{1}{\sqrt{3}}(1,-1,-1)   \!\!\! \quad . \label{eq:points1}
\end{eqnarray}
We now perform a stereographic projection (see Fig.
\ref{fig:stereo})
%%%%%%%%%%%%%%%%%%%%%%%%%%%%%%%%%
%%%%%%%%%%%%%%%%%%%%%%%%%%%%%%%%%%%%
that maps every point on a unit sphere centered on the origin onto the plane $z=-1$
according to the rule
\begin{eqnarray}
\left(%
\begin{array}{c}
  x \\
  y \\
  z \\
\end{array}%
\right) \rightarrow
\left(%
\begin{array}{c}
  a \\
  b \\
  -1 \\
\end{array}%
\right) \!\!\! \quad . \label{eq:points2}
\end{eqnarray}
The coordinates of the image points (connected to points on the sphere by dashed lines in Fig. \ref{fig:stereo}) are given by
\begin{eqnarray}
a&=&\frac{2 x}{1-z} \!\!\! \quad , \nonumber\\
b&=&\frac{2 y}{1-z} \!\!\! \quad . \label{eq:ab}
\end{eqnarray}
Upon transforming to a complex coordinate $w=a+ib$, the four tetrahedral points of Eq.(\ref{eq:points1})  are mapped onto
\begin{eqnarray}
A'&=& p \!\!\!\! \quad (1+i)   \!\!\! \quad , \!\! \quad
B'= p \!\!\!\! \quad (-1-i)   \!\!\! \quad , \nonumber\\
C'&=& q \!\!\!\! \quad (-1+i) \!\!\!\! \quad , \!\! \quad D'= q
\!\!\!\! \quad (1-i) \!\!\! \quad , \label{eq:points3}
\end{eqnarray}
where
\begin{eqnarray}
p=\sqrt{3}+1 \!\!\! \quad , \!\!\! \quad q=\sqrt{3}-1 \!\!\! \quad
. \label{eq:points4}
\end{eqnarray}
(In this section $p$ does not refer to the symmetry of the order parameter).
The great arc passing through the south pole (corresponding to one capacitor plate in the electrostatic analogy) maps onto the segment
$A'B'$, as illustrated schematically in the top panel of Fig. \ref{fig:mapping1}, while the great arc through the north pole
%%%%%%%%%%%%%%%%%%%%%%%%%%%%%%%%%%
%%%%%%%%%%%%%%%%%%%%%%%%%%%%%%%%%%
maps onto the two semi-infinite segments of the line $a+b=0$ which bracket
$C'D'$. We can fold the two cuts in the $w$ plane back on top of each
other by mapping the $w$ plane onto the $u$ plane via
\begin{eqnarray}
u=-i w^2 \!\!\! \quad . \label{eq:map1}
\end{eqnarray}
As shown in the middle panel of Fig. \ref{fig:mapping1}, the images
\~A and \~B  of A' and B' now both lie on the real axis at $2p^2$ while the 
images of C' and D' now lie at $-2q^2$. 
The two cuts in the $u$ plane are both on the real axis, running
from zero to $2p^2$ and from $-2q^2$ to minus infinity. On the sphere, these correspond to 
geodesics connecting defects which stretch more than halfway around the sphere. In order
to make the cuts symmetric with respect to the imaginary axis (see
the bottom panel of Fig. \ref{fig:mapping1}) we search for a conformal
transformation that maps the following four points in the complex $u$-plane to four points on 
the real axis of a complex $v$-plane
\begin{eqnarray}
u_{0}=0 & \rightarrow & v_{0}=k \!\!\! \quad , \nonumber\\
u_1=-2q^2 & \rightarrow & v_1=-k \!\!\! \quad , \nonumber\\
u_2=-\infty & \rightarrow & v_2=-1 \!\!\! \quad , \nonumber\\
u_3=2p^2 & \rightarrow & v_3=1 \!\!\! \quad .
\label{eq:points3}
\end{eqnarray}
In order to fully determine the conformal transformation we need to determine the value of $k$. This can be done by using a standard relation in the theory of conformal transformations \cite{Smirnov3.2}
\begin{eqnarray}
\frac{u_{0}-u_1}{u_{0}-u_2} \!\!\! \quad \frac{u_3-u_2}{u_3-u_1} =
\frac{v_{0}-v_1}{v_{0}-v_2} \!\!\! \quad \frac{v_3-v_2}{v_3-v_1}  \!\!\!
\quad . \label{eq:formula}
\end{eqnarray}
Upon inserting the points of
Eq.(\ref{eq:points3}) into Eq.(\ref{eq:formula}),
we determine the value of $k$ (less than one)
\begin{eqnarray}
k = \frac{2 \sqrt{2} - p}{p + 2 \sqrt{2}} \!\!\! \quad .
\label{eq:k}
\end{eqnarray}
Equation (\ref{eq:points3}) contains four independent relations so
we are still left with three conditions to determine the three
independent coefficients $\{\alpha, \!\!\!\!\! \quad \beta,
\!\!\!\!\! \quad \delta \}$ of the bilinear conformal transformation that
implements the mapping illustrated pictorially in the bottom plate
of Fig. \ref{fig:mapping1}
\begin{eqnarray}
v = \frac{u+\delta}{\alpha \!\!\!\!\! \quad u + \beta}  \!\!\!
\quad . \label{eq:bilinear1}
\end{eqnarray}
The required coefficients needed to implement the mapping in Eq.(\ref{eq:points3})
are  
\begin{eqnarray}
\alpha &=& -1 \!\!\! \quad , \nonumber\\
\beta &=& 2p \!\!\!\! \quad (2\sqrt{2}+p)  \!\!\! \quad , \nonumber\\
\delta &=& 2p \!\!\!\! \quad (2\sqrt{2}-p) \!\!\! \quad ,
\label{eq:parameters}
\end{eqnarray}
To solve Laplace's equation, we desire a function $\Omega(v)$ which is analytic except on the two cuts on the
real axis, and whose real part takes constant values on
the cuts. By symmetry, $\Omega(v)$ is an odd function of v, and
its real part $\Phi(v)$ is zero when the real part of v is zero.
Therefore the image of the seam in the v plane is simply the imaginary
axis $\Re e \!\!\!\! \quad v =0$. Upon substituting for $v$ using Equations
(\ref{eq:bilinear1}) and (\ref{eq:map1}), the condition $\Re e \!\!\!\! \quad v =0$
becomes
\begin{eqnarray}
16 + 4 p^2 \Im m(w^2)-|w|^4 = 0 \!\!\! \quad , \label{eq:seam1}
\end{eqnarray}
With the help of Equations (\ref{eq:ab}) and (\ref{eq:points2}), we can now
write down the equation of the seam explicitly in the original
cartesian coordinates \cite{Dyson}
\begin{eqnarray}
z = (2+\sqrt{3}) \!\!\!\!\! \quad xy \!\!\! \quad ,
\label{eq:seam1}
\end{eqnarray}
or in spherical polar coordinates $\{ \phi , \!\!\!\!\!\! \quad
\theta \}$ as
\begin{eqnarray}
\frac{\cos \theta}{\sin^2 \theta} =
\left(1+\frac{\sqrt{3}}{2}\right) \sin 2\phi \!\!\! \quad .
\label{eq:seam2}
\end{eqnarray}
The seam defined by the line of zero potential, is represented for different orientations of the sphere in
Fig. \ref{fig:seam3d}.
%%%%%%%%%%%%%%%%%%%%%%%%%%%%%%%%%%%%%
%%%%%%%%%%%%%%%%%%%%%%%%%%%%%%%%%%%%%
Its contour length $l$, on a unit radius ball, is readily calculated upon
integrating the expression for the infinitesimal arc of the seam
\begin{eqnarray}
d l = \sqrt{\sin ^2 \theta \left(\frac{d \phi}{d \theta}\right)^2
+1} \!\! \quad d\theta \!\!\! \quad , \label{eq:lineint}
\end{eqnarray}
from $\theta_{min}\approx 0.69$ radians to $\theta_{max}\approx
2.44$ radians and multiplying the result by four in view of the
symmetry of the seam. The values of $\theta_{min}$ and $\theta_{max}$ are obtained from Eq.(\ref{eq:seam2}) by setting $\phi$ equal to $\frac{\pi}{4}$ and $\frac{3 \pi}{4}$ respectively.  Upon using Eq.(\ref{eq:seam2}) to substitute
$\phi(\theta)$ in Eq.(\ref{eq:lineint}), we obtain $l \simeq 9.09$
for a sphere of unit radius. The seam is longer than the equatorial circumference
by slightly less than $50\%$.

\begin{figure}
\includegraphics[width=0.45\textwidth]{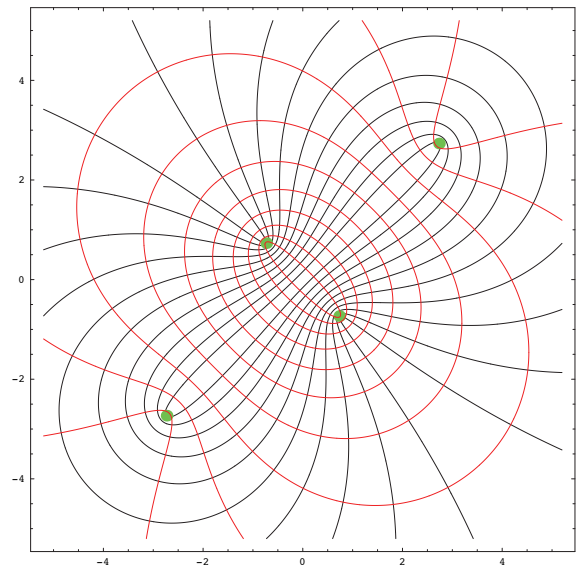}
\caption{\label{fig:nematic-plot} Illustration of the nematic
texture in stereographic projection (Color online). The red and black field
lines correspond to two energetically degenerate (in the one Frank
constant approximation) families of bend and splay rich textures.
The dots indicate the four tetrahedral $s=\frac{1}{2}$ disclination defects.}
\end{figure}

The branch cut structure in the $v$ plane is sufficiently simple
to allow a guess of the corresponding analytic function
$\Omega(v)$. A function with cuts from $k$ to $1$ and $-k$ to
$-1$, whose real part is equal and opposite on the two cuts and
with a single imaginary period around any curve separating the
cuts is easily identified to be a standard elliptic integral
\begin{eqnarray}
\Omega(v) = \int_{0}^{v} \left[(k^2-t^2)(1-t^2)\right]^{-
\frac{1}{2}} dt \!\!\! \quad . \label{eq:omega-nem}
\end{eqnarray}
with $v$ given in terms of $w$ by Equations (\ref{eq:map1}) and
(\ref{eq:bilinear1}). The nematic director is oriented (up to a
global rotation) along the contour lines of the imaginary or (real
part) of $\Omega (v)$. The equipotential (red) and field lines
(black) of $\Omega(v)$ are conveniently plotted using the
stereographic projection plane $w=a+ib$ in Fig.
\ref{fig:nematic-plot} along with the positions of the
disclinations (green dots).
%%%%%%%%%%%%%%%%%%%%%%%%%%%%
%%%%%%%%%%%%%%%%%%%%%%%%%%%%%%
It is easy to switch from the stereographic-projection plane
$w=a+ib$ of Fig. \ref{fig:nematic-plot} to spherical polar
coordinates $(\theta,\phi)$ by using the relation (reviewed in Appendix \ref{appB}), 
\begin{eqnarray}
w = 2 R \cot \left(\frac{\theta}{2} \right)e^{i\phi} \!\!\! \quad ,
\label{eq:stereo-coords}
\end{eqnarray}
where $R$ is the radius of the sphere.

If we had constructed the base-ball with cuts along the $short$
geodesics connecting the defects, then the form of the texture given in
Eq.(\ref{eq:omega-nem}) would be the same, but the parameter $k$ in
the elliptic integral would be given by
\begin{eqnarray}
k = (\sqrt{3}-\sqrt{2})^2 (\sqrt{2}+1)^2 \!\!\! \quad ,
\label{eq:stereo-coords}
\end{eqnarray}
instead of Eq.(\ref{eq:k}). The equation for the seam becomes
\begin{eqnarray}
z = -(2-\sqrt{3}) \!\!\!\! \quad  xy \!\!\! \quad
\label{eq:stereo-coords}
\end{eqnarray}
and the corresponding equipotential and field lines are plotted in
black and red respectively in Fig. \ref{fig:nematic-plot}. 
The expression reads 
\begin{eqnarray}
\frac{\cos \theta}{\sin^2 \theta} =-
\left(1-\frac{\sqrt{3}}{2}\right) \sin 2\phi \!\!\! \quad ,
\label{eq:seam3}
\end{eqnarray}
in spherical polar coordinates, and leads to the same contour length $l$ as Eq.(\ref{eq:seam2}).
Thus the two choices of arcs lead to two equivalent textures differing
only by a $\frac{\pi}{2}$ about the local normal. As in Section \ref{tilt}, the degeneracy in
energy between the red and black flow lines in Fig.
\ref{fig:nematic-plot} is lifted upon considering the effect of
elastic anisotropy generated by a different energy cost for bend
and splay. 

\section{Stability of liquid crystal textures to thermal fluctuations \label{sec:stability}}

In this section, we study the stability of liquid crystal ground
states to thermal fluctuations \cite{Nels02}. To explore the
fidelity of directional bonds at finite temperatures, we employ a Coulomb gas
representation of the liquid crystal free energy (in the one Frank
constant approximation) obtained by substituting in
Eq.(\ref{eq:patic-ener}) the relation
\begin{equation}
\gamma^{\alpha\beta}\partial_{\alpha}(\partial_{\beta}\theta-A_{\beta})
= s({\bf u})- G({\bf u}) \equiv n({\bf u}) \!\!\!\! \quad ,
\label{eq:curl-vel-d}
\end{equation}
where $\gamma^{\alpha\beta}$ is the covariant antisymmetric
tensor \cite{Davidreview}, $G({\bf u})$ is the Gaussian curvature and $s({\bf
u})\equiv\frac{1}{\sqrt{g}}\sum_{i=1}^{N_{d}}q_{i}\delta({\bf
u}-{\bf u}_{i})$ is the disclination density with $N_d$ defects of
charge $q_{i}$ at positions ${\bf u}_{i}$. The final result is an
effective free energy whose basic degrees of freedom are the
defects themselves \cite{Park96,bowi2000,Nels87}:
\begin{eqnarray}
F &=& \frac{K}{2}\int\! dA \!\!\!\!\! \quad \int \! dA'
\!\!\!\quad  n({\bf u}) \!\!\!\quad \Gamma({\bf u},{\bf u}')
\!\!\!\quad n({\bf u}') \!\!\!\! \quad . \label{Coulomb
model-energy-d}
\end{eqnarray}
The Green's function $\Gamma({\bf u},{\bf u}')$ is calculated (see
Appendix \ref{appA}) by inverting the Laplacian defined on the
sphere
\begin{eqnarray}
\Gamma({\bf u},{\bf u}')\equiv -
\left(\frac{1}{\Delta}\right)_{{\bf u}{\bf u}'} \!\!\!\! \quad ,
\label{invert}
\end{eqnarray}
and we have suppressed for now defect core energy contributions
which reflect the physics at microscopic length scales.
Equations (\ref{eq:curl-vel-d}) and (\ref{Coulomb
model-energy-d}) can be understood by analogy to
two dimensional electrostatics, with the Gaussian curvature
$G({\bf u})$ (with sign reversed) playing the role of a uniform
background charge distribution and the topological defects
appearing as point-like sources with electrostatic charges equal
to their topological charge $q_{i}$. The charge $q$ can
be defined by the amount $\theta$ increases along a counterclockwise path enclosing
the defect's core. On a generic surface, the defects tend to
position themselves so that the Gaussian curvature is screened:
typically, the positive ones are attracted to peaks and valleys
while the negative ones to the saddles of the surface
\cite{ViteNels04,ViteTurn05}. This geometric potential is ruled
out by symmetry on an undeformed sphere since the Gaussian
curvature is constant. The Gaussian curvature plays the role of a
uniform background charge fixing the net charge of the
defects consistent with the topological constraint imposed by the
Poincare-Hopf theorem (see section \ref{sec:texture} and
References \cite{Kami02,Needham-book}). The equilibrium positions
of the defects are then determined only by defect-defect
interactions which are proportional to the logarithm of their
chordal distance (see Appendix \ref{appA}) according to
\begin{equation}
F = -\frac{\pi K}{2 p^2} \sum_{i \neq j} n_{i} n_{j} \ln \left[1-
\cos \beta_{ij} \right] \!\!\!\! \quad , \label{eq:defectint1}
\end{equation}
where the integers $n_{i}$ and $n_{j}$ describing the
singularities associated with each defect, and the integer $p$
controls the period $\frac{2 \pi}{p}$ of the orientational order
parameter. The "topological charge" describing the rotation of the order parameter around 
each defect is given by $s_{j}=\frac{n}{p}$. 
The geodesic angle $\beta _{ij}$ subtended by the two defects
at positions $\bf{u}_{i}=\{\theta_{i},\phi_{i}\}$ and
$\bf{u}_{j}=\{\theta_{j},\phi_{j}\}$ can be conveniently recast in
terms of their spherical polar coordinates
\begin{equation}
\cos \beta _{ij} = \cos \theta_{i} \cos \theta_{j} + \sin
\theta_{i} \sin \theta_{j} \cos\left[\phi_{i}-\phi_{j}\right]
\!\!\!\! \quad . \label{eq:defectint2}
\end{equation}

As a simple example, we first consider the case of a $Z=2$
colloidal particle (p=1) with two antipodal defects of index $1$.
We can study the effect of thermal disruption of the ground state
by setting $\beta_{ij}=\pi + \theta$ in Eq.(\ref{eq:defectint1})
and expanding in the bending angle $\theta$. The resulting free
energy, apart from an additive constant, reads
\begin{equation}
F \approx \frac{\pi K}{4} \!\!\!\! \quad \theta^2  \!\!\!\!
\quad . \label{eq:frenergyvib1}
\end{equation}
Upon applying the equipartition theorem we obtain in the limit
$K \gg k_{B} T$ \cite{Nels02}
\begin{eqnarray}
\langle \cos \beta_{ij} \rangle &\approx& -1 + \frac{1}{2} \langle \theta ^ 2\rangle \!\!\!\! \quad , \nonumber \\
&\approx& -1 + \frac{k_{B}T}{\pi K} \!\!\!\! \quad ,
\label{eq:fluctcos}
\end{eqnarray}
which describes the fidelity of $\pi$ antipodal "bonds" of a divalent colloidal particle.

The effect of thermal fluctuations on the tetrahedral ground state of
nematic molecules confined on the sphere (p=2 and all
$s_{j}=\frac{1}{2}$) can be studied by means of a normal mode
analysis. The basic results sketched in \cite{Nels02} were obtained by
a slightly different method. Here we describe an alternative treatment in some detail 
and extend our analysis to hexatic and tetratic defect arrays (see Appendix \ref{appC}).  

We start by defining a generalized array of defect coordinates $\{q_{i}\}$ as
a $2N$ dimensional vector, where $N$ is the number of defects in
the ground state (or, equivalently, the valence of the colloidal
molecule). $N=4$ in the case of the tetrahedron. The first $N$
entries of the vector $\bf{q}$ are the longitudinal deviations of
the $N$ defects from a perfect tetrahedral configuration while the
remaining $N$ components describe defect
displacements along the lines of latitude of a sphere of unit
radius. As a result, the deviations of the $i^{th}$ defect from
its equilibrium configuration $\{\theta_{i} \!\!\!\!\!\!\! \quad
^{0},\phi_{i} \!\!\!\!\!\!\! \quad ^{0}\}$ are parameterized by
the two independent components of the vector $\bf{q}$
\begin{eqnarray}
q_i &=& \delta \theta_{i} \!\!\!\! \quad , \nonumber \\ q_{N+i}
&=& \delta \phi_{i} \!\!\!\! \quad \sin(\theta_{i}\!\!\!\!\!\!\!
\quad ^{0}) \!\!\!\! \quad . \label{eq:coords}
\end{eqnarray}
The relations in Eq.(\ref{eq:coords}) can be used to reexpress
Eq.(\ref{eq:defectint2}) in terms of the components of the
displacements vector $q_{i}$, with the result,
\begin{eqnarray}
\cos \beta _{ij} = \cos \left(\theta _{i} \!\!\!\!\!\!\! \quad
^{0}+ q_{i}\right) \cos \left(\theta_{j} \!\!\!\!\!\!\! \quad ^{0}
+ q_{j}\right) + \nonumber \\  \sin \left(\theta_{i} \!\!\!\!\!\!\! \quad ^{0} +
q_{i}\right) \sin \left(\theta _{j} \!\!\!\!\!\!\! \quad ^{0} +
q_{j}\right) \cos\left[\phi_{i} \!\!\!\!\!\!\! \quad ^{0}-
\phi_{j} \!\!\!\!\!\!\! \quad ^{0} + \frac{q_{N+i}}{\sin \theta
_{i} \!\!\!\!\!\!\! \quad ^{0}}- \frac{q_{N+j}}{\sin \theta _{j}
\!\!\!\!\!\!\! \quad ^{0}} \right] \!\!\!\! \quad .
\label{eq:defectint3}
\end{eqnarray}
Upon substituting Eq.(\ref{eq:defectint3}) in
Eq.(\ref{eq:defectint1}), the free energy $F$ can be expanded
around the equilibrium configuration to quadratic order in $q_{i}$
with the result (apart from an additive constant)
\begin{equation}
F \approx \frac{1}{2} \sum_{ij} \!\!\!\! \quad M_{ij} \!\!\!\! \quad q_{i}
\!\!\!\!\! \quad q_{j} \!\!\!\! \quad  \!\!\!\! \quad ,
\label{eq:fexp}
\end{equation}
where the matrix, $M_{ij}$, describing the deformation of the
tetrahedral molecule is naturally defined as
\begin{equation}
M_{ij} = \left[\frac{\partial ^{2} F}{\partial q_{i} \!\!\!\!\!
\quad \partial q_{j}}\right]_{q_{i} , q_{j} = 0} \!\!\! \quad .
\label{eq:fexp2}
\end{equation}
The eigenvalues of this matrix can be classified according to the
irreducible representation of the symmetry group of the
tetrahedron; their degeneracies can be determined purely from
the group theoretical relation \cite{WilsonDeciusCross-book,Landau-quantum-book}
\begin{equation}
n^{(\gamma)} = \frac{1}{g} \sum_{i} \!\!\!\!\! \quad g_{i}
\!\!\!\!\! \quad \chi_{i} ^{(\gamma) \ast} \!\!\!\!\! \quad
\chi_{i} ^{(\Sigma)}  \!\!\! \quad , \label{eq:group1}
\end{equation}
where $n^{(\gamma)}$ is the number of frequency degenerate normal
modes that transform like the irreducible representation labeled
by $\gamma$, $g_{i}$ is the number of symmetry operations of the tetrahedral point group in the
i$^{th}$ class, $g=\sum_{i} g_{i}=24$ is the total number of symmetry
operation in the group, $\chi_{i} ^{(\gamma)}$ is the character of
the i$^{th}$ class in the irreducible representation labelled by
$\gamma$ while $\chi_{i} ^{(\Sigma)}$ is the corresponding
character for the reducible representation formed by the defects'
displacements.

The information necessary to apply Eq.(\ref{eq:group1}) to a
tetravalent  colloid is collected in Table \ref{tab:tetrahedron}.
The top row contains the five symmetry class $\{E, C_3,
C_{2}, S_4, \sigma_{d}\}$ contained in the tetrahedral point group
$\Im _d$, corresponding respectively to the identity, three and two
fold rotations, four fold rotatory-reflections and reflection
through a plane of symmetry \cite{WilsonDeciusCross-book}. The
number of symmetry operations $g_{i}$ included in the i$^{th}$
class also appears in the top row: thus, $\{g_{i}\}=\{1,8,3,6,6\}$ where the same ordering used above to
list the classes has been adopted.
%%%%%%%%%%%%%%%%%%%%%%%%%%%%%%%%%%%%%
\begin{table}
\caption{\label{tab:tetrahedron} Character for the irreducible
representations of the tetrahedral point group together with the
character of the eight-dimensional representation $\Sigma$
generated by the defect displacements of a tetravalent colloid.}
\begin{ruledtabular}
\begin{tabular}{lccccr}
$\Im _d$ &  $E$ & 8$C_3$ & 3$C_{2}$ & 6$S_4$ & 6$\sigma_{d}$ \\
\hline
$A_1$ & 1 & 1 & 1 & 1 & 1 \\
$A_2$ & 1 & 1 & 1 &$-1$ & $-1$ \\
$E$ & 2 & $- 1$ & 2 & 0 & 0 \\
$F_1$ & 3 & 0 & $-1$ & 1 & $-1$ \\
$F_{2}$ & 3 & 0 & $-1$ & $-1$ & 1 \\
 & & &  &  &  \\
$\Sigma$ & 8 & $-1$ & 0 & 0 & 0 \\
\end{tabular}
\end{ruledtabular}
\end{table}
%%%%%%%%%%%%%%%%%%%%%%%%%%%%%%%%%%%%%
The left most column of Table \ref{tab:tetrahedron} lists the one,
two and three dimensional irreducible representations of the
tetrahedral group $\{A_1, A_2, E, F_1, F_2 \}$, along with the
eight dimensional representation $\Sigma$ generated by the defect
displacements. The entries of the table list the characters
corresponding to each class of the five irreducible
representations, $\chi_{i} ^{(\gamma)}$, and in the last row the
corresponding characters, $\chi_{i} ^{(\Sigma)}$, for the eight
dimensional representation. The former are tabulated from standard
group theoretical treatments while the latter needs to be worked
out from the traces of the transformation matrices that describe
how the displacement coordinates $q_{i}$ transform under the
action of each symmetry element in the group. These manipulations
are rather cumbersome, especially for the "icosahedral molecule"
arising when a spherical surface is coated with a pure hexatic
layer (see Appendix \ref{appC}).

In the rich literature on molecular vibrations a set of empirical
rules has been developed to write down the characters by examining
only the transformation of the $three$ dimensional cartesian
displacements of the few atoms whose equilibrium positions are not
altered by the symmetry operation. In Appendix \ref{appC} we
provide analogous rules that simplify the task of finding the
$\chi_{i} ^{(\Sigma)}$ characters by incorporating the constraint
that each atom is confined on a sphere and hence only $two$
orthogonal displacements need to be considered as shown in
Eq.(\ref{eq:coords}).

The interested reader is referred to Appendix \ref{appC} for a
more comprehensive mathematical justification of the normal mode
analysis applied to the tetrahedral colloid and to the more
complicated cases of hexatic $Z=12$ and tetratic order $Z=8$.
Here, we simply summarize the results of applying
Eq.(\ref{eq:group1}) in conjunction with Table
\ref{tab:tetrahedron} to find the degeneracies of the eigenvalue
spectrum of the matrix $M_{ij}$. The representation $\Sigma$
contains (only once) the three dimensional representations $F_2$
and $F_1$ as well as the two dimensional representation $E$.
\begin{equation}
\Sigma = F_2 + F_1 + E \!\!\! \quad . \label{eq:group2}
\end{equation}
The three normal coordinates with vanishing frequency correspond
to the three rigid body rotations and belong to the $F_1$
irreducible representation \cite{WilsonDeciusCross-book,Landau-quantum-book}. We are
left with a doublet ($E$) and a triplet ($F_{2}$) corresponding
respectively to two shear-like twisting deformations of the
tetrahedron and to three stretching and bending modes of the cords
joining neighboring defects.

This symmetry analysis is confirmed by direct diagonalization of
the matrix $M_{ij}$ which leads the following set of eigenvalues
$\lambda _{i}$
\begin{equation}
\{\lambda _{i} \}= \frac{3 \!\!\!\!\! \quad \pi \!\!\!\!\! \quad
K}{8} \!\!\! \quad \{0,0,0,1,1,2,2,2\} \!\!\! \quad .
\label{eq:group2}
\end{equation}
In Section \ref{appC}, we also list the eigenvectors $w_{i}$ of
$M_{ij}$. The displacement coordinates are readily expressed in
terms of the eigenvectors
\begin{equation}
q_i=U^{-1}_{ij} \!\!\!\!\! \quad w_{j} \!\!\! \quad ,
\label{eq:group3}
\end{equation}
where the unitary matrix $U$ diagonalizes $M$ and hence the free
energy of Eq.(\ref{eq:fexp}) and is defined by
\begin{equation}
U \!\!\!\!\! \quad M \!\!\!\!\! \quad U^{-1} = Diag \!\!\!\!\!
\quad (\lambda _{i}) \!\!\! \quad . \label{eq:group4}
\end{equation}
Its construction is easily achieved by the standard Gram-Schmidt
orthogonalization procedure to the eigenvectors
$\{w_{i}\}=\{w_1,...,w_8 \}$, where the same ordering chosen in
listing the eigenvalues in Eq.(\ref{eq:group2}) is implicitly
assumed. The resulting orthogonal basis vectors are the rows of
the $8 \times 8$ matrix $U$.

We are now in a position to evaluate $\langle \cos \beta _{ij}
\rangle$ where the thermal average is performed with the Boltzman
weight obtained from the free energy in Eq.(\ref{eq:fexp}) which
is now diagonal. Note that for the tetrahedron any choice of pair
of defects labelled by $i$ and $j$ (where $i \neq j$) will lead to
the same answer, unlike the less symmetric cases of the twisted
cube (p=4) and the icosahedron (p=6) considered in Appendix
\ref{appC}. The bending angle $\cos \beta _{ij}$ in
Eq.(\ref{eq:defectint3}) can be Taylor expanded in the $q_{i}$.
The resulting expression is rather cumbersome, but once the
displacements $\{q_{i}\}$ are reexpressed in terms of the normal
coordinates $\{w_{i}\}$ (by means of Eq.(\ref{eq:group3})), $\cos
\beta _{ij}$ reduces to
\begin{equation}
\cos \beta _{ij} = -\frac{1}{3} + \frac{2}{9} \left( w_{6} ^{2} +
w_{7} ^{2} + w_{8} ^{2} \right) \!\!\! \quad , \label{eq:group5}
\end{equation}
where the only eigenmodes $\{w_{6}, w_{7}, w_{8} \}$ appearing in
Eq.(\ref{eq:group5}) correspond to the bending triplet of
Eq.(\ref{eq:group2}).

It is now easy to perform the thermal average by Gaussian
integration of the energetically degenerate eigenmodes, with the result \cite{Nels02}
\begin{eqnarray}
\langle \cos \beta _{ij} \rangle &=& -\frac{1}{3} + \frac{2}{9} \!\!\!\!\! \quad \left( \langle w_{6} ^{2} \rangle +
\langle w_{7} ^{2} \rangle + \langle w_{8} ^{2} \rangle \right) \nonumber \\
&=& -\frac{1}{3} + \frac{16 \!\!\!\!\! \quad k_{B} \!\!\!\!\!\!
\quad T}{9 \pi K} \!\!\! \quad . \label{eq:result}
\end{eqnarray}

\section{\label{sec:valence} Valence transitions in thick nematic shells}

In this section, we study the crossover from a two to three
dimensional regime as the thickness of the spherical shell, $h$,
increases. For thicker shells, three dimensional defect
configurations ("escaped" in the third dimension) compete with the planar textures described in the
previous sections, leading to a structural transition and a change in valence from $Z=4$ to $Z=2$ beyond 
a critical value of $h$. 

We first consider the case of a cylindrical slab (or disk) of radius $R$ and
thickness $h$ filled with a nematic whose director is tangent to
the two circular faces \cite{Chic02} (see Fig.
\ref{fig:fieldslab}). This simpler geometry captures the essential
features of the problem and provides a suitable starting point for
understanding thin spherical shells (see Fig. \ref{fig:shell}).

\subsection{Slab geometry}

\begin{figure}
\includegraphics[width=0.4\textwidth]{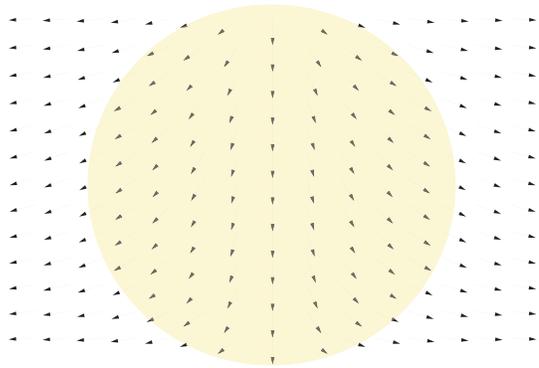}
\caption{\label{fig:fieldslab} Side view of the 2D nematic texture ansatz solution in
Eq.(\ref{eq:flow}). The cylindrical slab has height $h$ and radius $R$. The arrows can be interpreted as flow lines of a fluid entering a
narrow and long channel from a point source located on the top plate. To obtain the nematic texture the 
vectors need to be normalized and viewed as rods with both ends identified as shown in Fig. \ref{fig:shell}.}
\end{figure}

\begin{figure}
\includegraphics[width=0.35\textwidth]{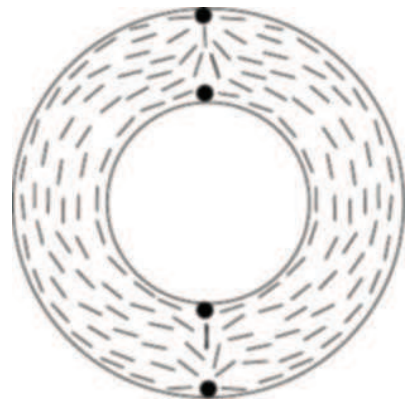}
\caption{\label{fig:shell} Nematic texture in a thin spherical
shell. The nematic director near the pair of half hedgehogs
indicated in the figure is well described by the one calculated
for the slab in Fig. \ref{fig:fieldslab}.}
\end{figure}

To estimate the energy stored in the texture of Fig. \ref{fig:fieldslab}, we coarse grain the
system to "blobs" of size $h$. The elastic energy arises from two
sources: a long distance contribution from a radial texture associated with an $s=1$ disclination and a local energy cost for the elastic deformations
inside the spherical blob in Fig. \ref{fig:fieldslab}.
%%%%%%%%%%%%%%%%%%%%%%%%%%%%%%%%
%%%%%%%%%%%%%%%%%%%%%%%%%%%%%%%%
In the one Frank constant approximation, the former can be estimated as the
energy $\pi K h \ln \left( \frac{R}{h} \right)$ of a disclination
whose enlarged "core" of size $h$ is given by the spherical blob while the latter
is roughly $4 \pi K h$, the energy of two half hedgehogs living
inside the blob \cite{Stark01}.
%%%%%%%%%%%%%%%%%%%%%%%%%%%%%%%%%%%%%
%%%%%%%%%%%%%%%%%%%%%%%%%%%%%%%%%%%%%%%
In view of the azimuthal symmetry of our configuration, the
director, $\textbf{n}(\textbf{r},z)$, can be parameterized by the angle
$\Theta(\textbf{r},z)$ formed with respect to the $\widehat{z}$ axis of the circular slab
along the centers of the two half hedgehogs shown in Fig. \ref{fig:fieldslab}
\begin{equation}
\textbf{n}(\textbf{r},z) = \sin \Theta (\textbf{r},z) \textbf{e}_r + \cos \Theta (\textbf{r},z)
\textbf{e}_z \!\!\!\! \quad . \label{eq:spin-wave}
\end{equation}
The energy density $f(\Theta)=\frac{1}{2}K_{1}\left(\nabla \cdot \textbf{n} \right)^{2}+\frac{1}{2}K_{3}\left(\nabla \times \textbf{n} \right)^{2}$ expressed in terms of the bond
angle reads
\begin{eqnarray}
f(\Theta) &=& \frac{K_1}{2} \left(\frac{\sin \Theta}{r} + \Theta_r
\cos \Theta - \Theta_z \sin \Theta \right)^{2} \nonumber\\ &+&
\frac{K_3}{2} \left(\Theta_z \cos \Theta + \Theta_r \sin \Theta
\right)^{2} \!\!\!\! \quad . \label{eq:free-energy}
\end{eqnarray}
where $K_{1}$ and $K_{3}$ are the splay and bend constants and $\Theta_{r}\equiv \frac{\partial \Theta}{\partial  r}$ and $\Theta_{z}\equiv \frac{\partial \Theta}{\partial z}$ . In the
one Frank constant approximation, minimization of the free energy leads
to a non linear partial differential equation for the bond angle
$\Theta(\textbf{r},z)$
\begin{eqnarray}
\frac{1}{r}\frac{\partial}{\partial r}\left(r \frac{\partial \Theta}{\partial r} \right)+\frac{\partial ^2 \Theta}{\partial z^{2} } = \frac{\sin 2 \Theta}{r^2} \!\!\!\! \quad .
\label{eq:PDE}
\end{eqnarray}
The operator on the left arise from the Laplacian in cylindrical coordinates.
Note there is no need to explicitly consider the
azimuthal angle $\phi$ as an additional independent variable in
view of the symmetry of the problem.

Instead of solving this partial differential equation, we follow a route analogous to that
in Ref.\cite{Lube98}, that is we construct an $exact$ 2D solution
for the liquid crystal problem and then rotate it along the axis
$\widehat{z}$ to retrieve an $ansatz$ for the 3D director
configuration. The 2D solution is the true
minimum of the two dimensional liquid crystal elastic free energy
while the 3D Ansatz does $\bf{not}$ minimize $f(\Theta)$ nor
satisfy Eq.(\ref{eq:PDE}).

To solve the 2D problem we adopt the method of conformal mappings
that simplifies the study of many complicated boundary problems in
fluid dynamics and electrostatics \cite{Smirnov3.2}. Think of Fig.
\ref{fig:fieldslab} as a source of fluid confined to flow in a
narrow and long channel ($R \gg h$). The spherical "half-hedgehog" corresponds to the
source and the hyperbolic one at the top to the stagnation point of the flow.
The complex potential $\Omega(w)$ of the desired flow is
\begin{eqnarray}
\Omega(w) = Log \left(\sin \left(\frac{\pi w}{h}\right)-1\right)
\!\!\!\! \quad ,  \label{eq:flow}
\end{eqnarray}
where the complex variable is denoted by $w=r+iz$ to distinguish it from 
the $z$ coordinate along the axis of the cylinder.
The velocity field is given by $\overline{\Omega'(w)}$ and the
corresponding complex nematic director $\textbf{n}(w)=\cos{\Theta}+i \sin{\Theta}$ 
is obtained by normalizing this
vector field. Note that the prime in $\Omega'(w)$ denotes the
derivative with respect to $w$ of the complex function $\Omega=\Phi+i\Psi$
and we define $\overline{\Omega}\equiv \Phi- i\Psi$. A straightforward but
tedious calculation (see Appendix B)
leads the functional form of $\Theta(r,z)$ for
a source at $z=-\frac{h}{2}$ and a stagnation point at
$z=+\frac{h}{2}$, namely 
\begin{eqnarray}
\tan \Theta (r,z) = \sec \left(\frac{\pi z}{h}\right)
\sinh\left(\frac{\pi r}{h}\right) \!\!\!\! \quad . \label{eq:flow}
\end{eqnarray}
This trial solution respects the boundary conditions of the
problem, has the correct short and long distance behavior in
$r$ and the expected functional form near the source.

Upon inserting $\Theta(r,z)$ in Eq.(\ref{eq:free-energy}) and
integrating the energy density $f(\Theta)$ over the cylindrical
volume of the slab, we obtain the total elastic energy $E_{1}$
stored in the field. The resulting energy $E_{1}$ does
$\bf{not}$ rely on the assumption $R\gg h$, as can be explicitly
checked numerically.
\begin{eqnarray}
E_{1} = \pi K h \left[\ln \left(\frac{R}{h}\right)+ c \right]
\!\!\!\! \quad , \label{eq:totalenergy}
\end{eqnarray}
where $c \approx 4.2$. Note that the functional
dependence of $E_{1}$ on $R$ and $h$ matches the expectations from
the blob argument. Indeed the prefactor of the logarithm is
universal in the sense that it does not depend on the details of
the trial solution near the hedgehogs, but only on its long
distance behavior. By contrast, we expect the result quoted above
for the coefficient $c$ to be only an estimate (an upper bound) since
it relies on a $trial$ solution for $\Theta(r,z)$ that is not the
absolute minimum of the free energy. Numerical studies carried out in Ref.\cite{Chic02} for the slab geometry reported that $c\approx4.19$.

A competing energy minimum for the nematic is given by a planar
texture with two $s=\frac{1}{2}$ disclination lines. Note that
$+\frac{1}{2}$ lines cannot escape in the third dimension
\cite{chandra-book}. A tedious but straightforward calculation
shows that the energy $E_{p}$ of the pair of disclination lines is
given by
\begin{eqnarray}
E_{p} = \frac{\pi}{2} K h \left(\ln \left(\frac{R}{a}\right) -
0.06 + \frac{4 \!\!\!\! \quad E_c}{\pi K h} \right) \!\!\!\! \quad
. \label{eq:pairenergy}
\end{eqnarray}
where $a$ is a macroscopic cutoff typically of the order of the molecular
length. These two disclination repel each other, and are repelled by the circular boundary, leading to a separation of order $R$. 
The second term of Eq.(\ref{eq:pairenergy}), corresponding to the interaction of the
two disclinations with the boundary and among themselves, is
negligibly small. The third term accounts for the core energies of
the two disclination lines $2E_c$. The combination $K h$
is equal to the two dimensional coupling constant $K_{2D}$. The relevant dimensionless
ratio is $\frac{E_c}{K_{2D}}$.

By setting $E_{p}=E_{1}$ we obtain the critical thickness $h^*$
above which the escaped "half-hedgehogs" become energetically
favored compared to a single $s=\frac{1}{2}$ disclination line:
\begin{eqnarray}
h^* = e^{c'} \sqrt{R \!\!\!\!\! \quad  a}  \!\! \quad .
\label{eq:hc1}
\end{eqnarray}
The core energy terms in Eq.(\ref{eq:pairenergy})
reduce the previous estimate of $c$ to $c'=c-\frac{2 E_{c}}{\pi K h}$. A similar analysis
applies to spherical shells which we now discuss.

\subsection{Extrapolation to thin spherical shells}

In principle, one could proceed along the same route as in the previous section
and find a conformal mapping that provides a trial function for
the bond angle $\Theta(\theta , r)$ corresponding to the texture shown in Fig. \ref{fig:shell}.
This is possible but rather cumbersome. In the case of very thin
shells one can adapt the slab calculation by
noting again that the energy is composed of two parts. There is
a long distance piece arising from "combing the hair" of the nematic
texture in the tangent plane of the sphere that we can read off from a suitable 2D calculation (see
Appendix \ref{appA}) and the short distance contribution arising
from the short distance contribution arising from the two pairs of half-hedgehogs at the north and south pole.

The energy $E_d$ of two short $+1$ disclination lines placed at antipodal
points on a sphere (the north and south pole, say) can be estimated by performing a 2D calculation on
the curved surface and simply multiplying the result by the
thickness of the layer $h$
\begin{eqnarray} E_{d} = 2 \pi K h \left(\ln
\left(\frac{R}{a}\right) - 0.3 +\frac{E_{c}}{\pi K h}\right) \!\!\!\! \quad ,
\label{eq:energy2unstable}
\end{eqnarray}
where the middle term accounts for the interaction between the two
disclinations in their equilibrium positions. Note that this result is accurate only up to
factors of the order of $\left(\frac{h}{R}\right)$ since the explicit
integration over the volume of the thin shell was bypassed.
To obtain the energy of the escaped solution,
the core size $a$ in Eq.(\ref{eq:energy2unstable}) is rescaled to
$h$. This will account for the integration of the energy density
at distances of the order of a few $h \ll R$ from the two
hedgehogs. In these portions of the shell the integrand
reduces to the energy density of the two disclination problem and
hence the integration can be easily carried out leading the result
in Eq.(\ref{eq:energy2unstable}) with a lower cutoff of the order
$h$.

The energy stored in the remaining portions of the thin shell is
approximately given by twice the energy $4.2 \!\!\!\!\! \quad \pi
K h$ of the yellow blob of Fig. \ref{fig:fieldslab}. This estimate
neglects curvature corrections of the order of
$\left(\frac{h}{R}\right)$ and arises because at
distances of the order $h$ the spherical shell looks $locally$
like a flat circular slab as long as $h \ll R$. The resulting
energy $E_2$ of the escaped configuration reads
\begin{eqnarray}
E_{2} = 2 \pi K h \left(\ln \left(\frac{R}{h}\right) - 0.3 + 4.2
\right) \!\!\!\! \quad . \label{eq:totalenergy}
\end{eqnarray}
Although the prefactor of the sub-leading term
linear in $h$ has only been estimated, we expect that the coefficients of the
logarithm, which arises from large scales compared to $h$,  is exact. For a
spherical shell whose radius $R$ is a hundred times its thickness,
the corrections from higher powers of $\frac{h}{R}$ are indeed negligible.
However for reasonable values of $\frac{R}{h}$, the logarithmic term of
Eq.(\ref{eq:totalenergy}) is still comparable in magnitude to the
"sub-leading" one linear in $h$.

The energy $E_4$ of the tetravalent configuration can be evaluated
using similar considerations, with the result
\begin{eqnarray}
E_{4} = \pi K h \left(\ln \left(\frac{R}{a}\right) - 0.4 + \frac{4 E_{c}}{K \pi h} \right)
\!\!\!\! \quad . \label{eq:e4}
\end{eqnarray}
Upon setting $E_4 = E_2$ we obtain the critical thickness $h^*$
below which the tetravalent configuration becomes energetically
favored
\begin{eqnarray}
h^* = e^{(4.1-\frac{2 E_{c}}{K \pi h})} \sqrt{R \!\!\!\!\! \quad  a}  \!\! \quad .
\label{eq:hc2}
\end{eqnarray}
The exponential prefactor arises from the terms linear in $h$ in
Eq.(\ref{eq:totalenergy}), which cannot be ignored in estimating
$h^*$ even in the limit $R \gg h$. Note that an accurate
determination of the argument in the exponent would require
knowledge of the core energies of the disclination lines. In fact, the exponential prefactor
can be interpreted as a numerically significant rescaling of the
core radius.

The energy barrier between these two coexisting minima of the free
energy $f(\Theta)$ can be estimated by splitting the path connecting
them in $\Theta$ space in two steps. First, consider a continuous
deformation of the escaped texture of Fig. \ref{fig:shell} obtained by appropriately rotating each nematigen
until the $non-escaped$ solution (with two disclinations $lines$
of index one at the north and south pole) is recovered. This part
of the path must be uphill in energy if the escaped solution was
allowed to escape in the first place. The corresponding energy
barrier $\Delta E$ is given approximately by the difference
between $E_d$ as calculated in Eq.(\ref{eq:energy2unstable}) and
$E_{2}$ in Eq.(\ref{eq:totalenergy})
\begin{eqnarray}
\Delta E = 2 \pi K h \left(\ln \left(\frac{h}{a}\right) -
4.2\right) \!\!\!\! \quad . \label{eq:barrierenergy}
\end{eqnarray}

The second step consists in letting each of the unstable
disclination lines split in two $+\frac{1}{2}$ defects and
subsequently separate them until they sit at the vertexes of a
tetrahedron inscribed in the sphere. This portion of the path is
downhill because the "non-escaped" texture of valence $2$ is
unstable. This can be proved by writing down the energy
of the pair and show that it decreases monotonically as one
separates them because of the "electrostatic-like" repulsion
\cite{lube92,ViteNels04}. As a result the energy barrier is
simply the energy difference calculated in
Eq.(\ref{eq:barrierenergy}).

Upon inserting $K \approx 10^{-6}$ dyn in
Eq.(\ref{eq:barrierenergy}) and taking $k_{B} T \approx 4 \!\!\!\!
\quad 10^{-14}$ erg, as in Ref.\cite{Poul97,Stark01}, we obtain
\begin{eqnarray}
\frac{\Delta E}{k_{B}T} \approx \frac{15 h}{nm} \left(\ln
\left(\frac{h}{a}\right) - 4.2 + \frac{E_{c}}{K \pi h}\right) \!\!\!\! \quad .
\label{eq:barriernumbers1}
\end{eqnarray}
For shells with critical thickness $h^*$
Eq.(\ref{eq:barriernumbers1}) reduces to
\begin{eqnarray}
\frac{\Delta E}{k_{B}T} \approx 10^3 \sqrt{\frac{R}{a}} \!\!\!\!
\quad \ln \left( \frac{R}{a}\right) \!\!\!\! \quad ,
\label{eq:barriernumbers2}
\end{eqnarray}
where a core size of the order of $10$ nm was assumed and the core 
energies were set to zero \cite{lube92}. This estimate indicates that the energy barrier is
very high around $h^*$ suggesting that exchange between the
two minima is unlikely to happen by thermal activation. In a
monodisperse solution of shells with thickness $h$, the ratio
between shells of the two valence will be given by their Boltzman
factors as long as equilibrium is reached. If one engineers shells
with thickness below $h^*$, the Z=4 configuration would be more
likely.

\section{\label{sec:conc}Conclusion}

We have studied the crossover from the two dimensional regime of liquid crystals confined on a spherical surface to the full three dimensional problem in a spherical shell. For very thin shells, the nematic ground state has four disclination lines sitting at the vertices of a tetrahedron inscribed in the ball and whose texture approximately track the seam of a baseball. As the thickness increases, a competing three dimensional defected texture characterized by two pairs of half hedgehogs at the north and south pole becomes energetically favorable. For ultra-thin shells this instability is suppressed and one expects a defected ground state with tetravalent symmetry. Estimates of the stability of this texture to thermal fluctuations indicate      
that the vibrations around the equilibrium configurations of the defects should not be significant. The present analysis has been carried out primarily in the limit in which the elastic anisotropy parameter, $\epsilon=\frac{K_{3}-K_{1}}{K_{3}+K_{1}} \ll 1$. 

We hope to extend our investigation with a systematic study of the effect of elastic anisotropy on the nematic texture. It is interesting to note that in the case of pure bend or splay (ie. $\epsilon = \pm 1$)  the ground state is given by only two disclinations of unit index at the north and south pole. This suggests the possibility that the effect of the elastic anisotropy may not be limited to locally adjusting the orientation of the director but may induce a change in the inter-defect interaction and hence a distortion of the tetravalent equilibrium configuration. The limit of strong elastic anisotropy is also relevant to studies of the nematic to smectic transition in a spherical geometry for which the ratio of the bend to splay coupling constants $\frac{K_{3}}{K_{1}}$  is expected to diverge.  

Additional experimental complications include the possibility of having a nematic layer of non constant thickness that would induce trapping of the defects in the regions where the layer is thinner. This effect may also induce a local transition to an escaped texture where the layer thickens in just one hemisphere so that two disclination lines of index $\frac{1}{2}$ are traded for a pair of half hedgehogs. If that happens shells with three-fold symmetry could be observed. 

\begin{acknowledgments}
We wish to acknowledge helpful conversations with A. Fernandez-Nieves, O. D. Lavrentovich, D. Link, P. J. Lu, A. Polkovnikov, A. M. Turner and D. Weitz. We are grateful 
to F. Dyson for the exact solution of the nematic texture.
This work was supported by the National
Science Foundation, primarily through the Harvard Materials
Research Science and Engineering Laboratory via Grant No.
DMR-0213805 and through Grant No. DMR-0231631.
\end{acknowledgments}

\appendix

\section{\label{appA} Free energy of a vector field on a sphere}

We start our analysis with the Frank free energy with splay and bend terms proportional to $K_{1}$ and $K_{3}$ and expressed in terms of the
covariant derivative $D_i \!\!\!\!\! \quad n^j$
\begin{eqnarray}
F&=&\frac{1}{2} \int d^2 \textbf{x} \!\!\!\! \quad \sqrt{g}
\!\!\!\!
\quad [K_1 \left(D_i \!\!\!\!\! \quad n^i \right)^2 \! \nonumber\\
&+& K_3 \left(D_i \!\!\!\!\! \quad n_j - D_j \!\!\!\!\! \quad n_i
\right)
 \left( D^i \!\!\!\!\! \quad n^j - D^j \!\!\!\!\!
\quad n^i \right)] \!\!\!\! \quad , \label{eq:frank-free}
\end{eqnarray}
where
\begin{equation}
D_i \!\!\!\!\! \quad n^j = \partial _{i} \!\!\!\!\! \quad n^j +
\Gamma \!\!\!\!\! \quad  ^{j}_{i \!\!\!\!\!\!\! \quad k} \!\!\!\!
\quad n^k \!\!\!\!\! \quad . \label{eq:frank1bis}
\end{equation}
The Christoffel connection $\Gamma \!\!\!\!\! \quad  ^{j}_{i
\!\!\!\!\!\!\! \quad k} $ \cite{Weinberg-GRbook} is unchanged if
the lower indices $i$ and $k$ are interchanged. As a result, the
covariant derivative $D_{i}$ can be replaced by $\partial _{i}$ in
the second term of Eq.(\ref{eq:frank-free}) because the covariant
curl is antisymmetric. It follows that
\begin{equation}
\vec{D} \times \vec{n} = \nabla \times \vec{n} \!\!\!\!\! \quad . \label{eq:curl}
\end{equation}
The covariant form of the divergence
is given by \cite{Weinberg-GRbook} 
\begin{equation}
D_i \!\!\!\!\! \quad n^i \equiv \frac{1}{\sqrt{g}} \!\!\!\! \quad
\partial _i \left(\sqrt{g} \!\!\!\!\! \quad n^i \right) \!\!\!\!
\quad . \label{eq:frank2bis}
\end{equation}
For a rigid sphere of radius $R$ with polar coordinates $\{
\theta,\phi \}$, we have $\sqrt{g}=R^2 \sin \theta $, $\Gamma
\!\!\!\!\! \quad  ^{\theta}_{\phi \!\!\!\!\!\!\! \quad
\phi}=-\sin\theta \cos\theta$, $\Gamma \!\!\!\!\! \quad
^{\phi}_{\phi \!\!\!\!\!\!\! \quad \theta}= \Gamma \!\!\!\!\!
\quad ^{\phi}_{\theta \!\!\!\!\!\!\! \quad \phi}= - \cot \theta$,
and all other $\Gamma \!\!\!\!\! \quad  ^{j}_{i \!\!\!\!\!\!\!
\quad k}=0$.

Upon adding and subtracting the expression for the curl of
$\bf{n}$ multiplied by $K_{1}$ from the first and second term in
Eq.(\ref{eq:frank-free}) we obtain
\begin{eqnarray}
F=\frac{1}{2} \int d^2 \textbf{x} \!\!\!\! \quad \sqrt{g} \!\!\!\!
\quad [K_1 \left(D_i \!\!\!\!\! \quad n^j\left) \!\!\!\! \quad
\right( D^i \!\!\!\!\! \quad n_j \right) \nonumber \\ + (K_3- K_1)
(\nabla \times \textbf{n})^2 ] \!\!\!\! \quad . \label{eq:frank1b}
\end{eqnarray}

Similarly, upon adding and subtracting the covariant divergence of
$\bf{n}$ multiplied by $K_{3}$ from the second and first term in
Eq.(\ref{eq:frank-free})  we obtain
\begin{eqnarray}
F=\frac{1}{2} \int d^2 \textbf{x} \!\!\!\! \quad \sqrt{g} \!\!\!\!
\quad [K_3 \left(D_i \!\!\!\!\! \quad n^j\left) \!\!\!\! \quad
\right( D^i \!\!\!\!\! \quad n_j \right) \nonumber\\
+ \left(K_1- K_3 \right) (\nabla \cdot \textbf{n})^2 ] \!\!\!\!
\quad . \label{eq:frank3b}
\end{eqnarray}
Upon adding the two equivalent expressions for $F$ in Equations
(\ref{eq:frank1b}) and (\ref{eq:frank3b}) and dividing by two we
can express the free energy in terms of the constants
\begin{eqnarray}
\epsilon &\equiv& \frac{K_3 - K_1}{K_3 + K_1}  \!\!\!\! \quad , \\
K &\equiv& \frac{K_3 + K_1}{2}  \!\!\!\! \quad ,
\label{eq:epsilonb}
\end{eqnarray}
namely
\begin{eqnarray}
F=\frac{K}{2} \int d^2 \textbf{x} \!\!\!\! \quad \sqrt{g} \!\!\!\!
\quad [  \left(D_i \!\!\!\!\! \quad n^j\left) \!\!\!\! \quad
\right( D^i \!\!\!\!\! \quad n_j \right) \nonumber \\ + \epsilon
\left( (\nabla \times \textbf{n})^2- (\nabla \cdot \textbf{n})^2 \right) ] \!\!\!\! \quad .
\label{eq:frankintb}
\end{eqnarray}

We now parameterize the orientation of the unit vector
$\bf{n}$ in terms of the bond angle field $\alpha(\theta,\phi)$
that it forms with respect to the longitudinal direction $\vec{e}$$_{\theta}$ which in
polar coordinates $(\theta,\phi)$ is given by
\begin{eqnarray}
\textbf{e}_{\theta} = R (\cos{\theta} \sin{\phi},\cos{\theta}\cos{\phi},-\sin{\theta})  \!\!\!\! \quad ,
\label{eq:etheta}
\end{eqnarray}
while the orthogonal unit vector $\textbf{e}_{\phi}$ is given by
\begin{eqnarray}
\textbf{e}_{\phi} = R (-\sin{\theta} \sin{\phi},\sin{\theta}\cos{\phi},0)  \!\!\!\! \quad .
\label{eq:etheta}
\end{eqnarray}
The components of the vector $\bf{n}$ with respect to $\vec{e}_{\phi}$ and $\vec{e}$$_{\theta}$ are given by
\begin{eqnarray}
n^{\theta} &=& \frac{\cos \alpha}{R}  \!\!\!\! \quad , \\
n^{\phi} &=& \frac{\sin \alpha}{R \sin \theta}  \!\!\!\! \quad .
\label{eq:vec}
\end{eqnarray}
Upon substituting the relevant expressions for the non vanishing
components of the connection in the covariant derivative (see
Eq.(\ref{eq:frank1bis})) and using Eq.(\ref{eq:vec}), the free
energy density $f(\theta,\phi)$ is given by
\begin{eqnarray}
\frac{4 R^2}{K_1 + K_3} \!\!\!\! \quad f &=& \left(\frac{\partial
\alpha}{\partial \theta}\right)^{2}  + \Upsilon _{\alpha}
\!\!\!\!\!\!\! \quad ^{2} (\theta,\phi) \nonumber\\ &+& \epsilon
\cos 2 \alpha \left[\left(\frac{\partial \alpha}{\partial
\theta}\right)^{2}- \Upsilon _{\alpha} \!\!\!\!\!\!\! \quad ^{2}
(\theta,\phi)\right] \nonumber\\ &+& 2 \epsilon \sin 2 \alpha
\left(\frac{\partial \alpha}{\partial \theta}\right) \Upsilon
_{\alpha} (\theta,\phi)
 \!\!\!\! \quad , \label{eq:frankexplicit}
\end{eqnarray}
where the $\alpha$-dependent function $\Upsilon_{\alpha}(\theta, \phi)$ is
\begin{eqnarray}
\Upsilon _{\alpha} (\theta,\phi)\equiv\frac{1}{\sin \theta}
\left(\frac{\partial \alpha}{\partial \phi} + \cos \theta
\right)\!\!\!\! \quad . \label{eq:frankexplicit2}
\end{eqnarray}
The energies of the latitudinal ($\alpha=\pi/2$) and longitudinal ($\alpha=0$) tilted-molecules textures favored for $\epsilon$ less or greater than zero respectively (see section \ref{tilt}) are easily determined by substituting the appropriate $\alpha$ and $\epsilon$ in Eq.(\ref{eq:frankexplicit}). After integrating
the free energy density $f$ in Eq.(\ref{eq:frankexplicit}) we obtain 
\begin{eqnarray}
F &=& 2 \pi K (1-|\epsilon|) \int_{\frac{a}{R}}^{\frac{\pi}{2}} \frac{\cos^2(\theta)}{\sin(\theta)} \!\!\!\!\! \quad d\theta \nonumber\\ &=& 2 \pi K (1-|\epsilon|) \left(\ln \left(\frac{2R}{a}\right)-1\right) + 2E_{c}
 \!\!\!\! \quad , \label{eq:frankexplicit3}
\end{eqnarray}
where we have introduced a core radius, $a$, and corresponding core energy $E_{c}$ for each defect. 

In the zero anisotropy limit ($\epsilon=0$), only the first line in Eq.(\ref{eq:frankexplicit2})
survives. The resulting free
energy $F=\int dS f(\theta,\phi)$ then matches the one obtained using the spin connection in the one Frank constant
approximation upon setting $K_{1}=K_{3}=K$,
\begin{equation}
F = \frac{1}{2}K\int dS \!\!\!\!\! \quad g^{ij}
(\partial_{i}\alpha - A_{i})(\partial_{j}\alpha - A_{j})\!\!\!\!
\quad , \label{eq:patic-enerbis}
\end{equation}
where $dS=d\theta d\phi \!\!\!\!\! \quad R^2 \sin \theta $ and the
metric tensor $g^{\phi\phi}=diag\left(\frac{1}{R^2},\frac{1}{R^2 \sin ^2 \theta}\right)$. The curl of the
spin-connection $A_{i}$ is the Gaussian curvature $\frac{1}{R^2}$
\cite{Davidreview,Kami02} and its only non-vanishing component is
$A_{\phi}=-\cos \theta$.

We now adopt a Coulomb gas representation of the liquid crystal
free energy (in the one Frank constant approximation) obtained by
exploiting in Eq.(\ref{eq:patic-enerbis}) the relation
\begin{equation}
\gamma^{ij}\partial_{i}(\partial_{j}\alpha-A_{j})
= s({\bf u})- G({\bf u}) \equiv n({\bf u}) \!\!\!\! \quad ,
\label{eq:curl-vel}
\end{equation}
where $\gamma^{ij}$ is the covariant antisymmetric
tensor \cite{Davidreview}, $G({\bf u})$ is the Gaussian curvature and $s({\bf
u})\equiv\frac{1}{\sqrt{g}}\sum_{i=1}^{N_{d}}q_{i}\delta({\bf
u}-{\bf u}_{i})$ is the disclination density with $N_d$ defects of
charge $q_{i}$ at positions ${\bf u}_{i}$. The final result is an
effective free energy whose basic degrees of freedom are the
defect positions themselves \cite{Park96,bowi2000}:
\begin{eqnarray}
F &=& \frac{K}{2}\int\! dA \!\!\!\!\! \quad \int \! dA'
\!\!\!\quad  n({\bf u}) \!\!\!\!\!\quad \Gamma({\bf u},{\bf u}')
\!\!\!\!\! \quad n({\bf u}') \!\!\!\! \quad , \label{Coulomb
model-energy}
\end{eqnarray}
where $n({\bf u})$, the defect density relative to the Gaussian curvature, was defined in Eq.(\ref{eq:curl-vel}). The
equilibrium positions of the defects are determined only
by defect-defect interactions because the Gaussian curvature is
constant $G=\frac{1}{R^2}$ on an undeformed sphere. To calculate the Green's
function $\Gamma({\bf u},{\bf u}')$ we need to invert the
covariant Laplacian defined on the sphere
\begin{eqnarray}
\Gamma({\bf u},{\bf u}')\equiv -
\left(\frac{1}{\Delta}\right)_{{\bf u}{\bf u}'} \!\!\!\! \quad ,
\label{invert}
\end{eqnarray}
As shown below, this inversion can be accomplished by performing a weighted sum over
eigenmodes of the covariant Laplacian, \cite{bowi2000}.

We first recall that the (generalized) Green function $\Gamma({\bf
u},{\bf u}')$ is defined by
\begin{eqnarray}
\Delta_{\textbf{u}} \!\!\!\!\! \quad \Gamma({\bf u},{\bf u}')=
\frac{\delta(\bf{u,u'})}{\sqrt{g}} - \frac{1}{S} \!\!\!\! \quad ,
\label{eq:defgreen}
\end{eqnarray}
where $S=4\pi R^2$ denotes the area of the surface and $\Delta=\frac{1}{\sqrt{g}}\partial_{i}(\sqrt{g} g^{ij} \partial_{j})$. The presence
of the second term on the left hand side of Eq.(\ref{eq:defgreen})
can be understood as follows. The Green
function of the Laplacian (according to the usual definition
without the area correction in Eq.(\ref{eq:defgreen})) can be
interpreted physically as the steady temperature response of the
system to a point-like unit source of heat. However, for a closed
system such as the surface of the sphere heat cannot escape.
Hence, it is impossible to impose a point source, that would
inject heat at a constant rate and have the system respond
with a time-independent distribution. To prevent energy from building up in such
a system, we put the spherical surface of area $S$ in contact with a reservoir
that uniformly absorbs heat at the same rate it is pumped in. The need for
subtracting the "neutralizing background heat" $\frac{1}{S}$ in Eq.(\ref{eq:defgreen}) will
become transparent mathematically once we proceed to determine
$\Gamma({\bf u},{\bf u}')$ explicitly.

The first step consists in writing the delta function as a sum
over spherical harmonics $Y^{m} _{l}(\theta, \phi)$,
\begin{eqnarray}
\frac{\delta (u,u')}{\sqrt{g}} &\equiv& \frac{\delta(\theta - \theta ') \delta(\phi
- \phi ')}{R^2 \sin \theta '} \nonumber\\ &=& \frac{1}{R^2} \sum
_{l=0} ^{\infty} \sum _{m=-l} ^{m=+l} Y^{m}_{l}(\theta,\phi)
{Y^{m} _{l}} ^{\ast}  (\theta,\phi) \!\!\!\!\! \quad ,
\label{eq:defdelta}
\end{eqnarray}
and recall the eigenvalue equation
\begin{eqnarray}
\Delta \!\!\!\!\! \quad Y^{m} _{l}  (\theta,\phi)= -
\frac{l(l+1)}{R^2} \!\!\! \quad Y^{m} _{l} (\theta,\phi)
\!\!\!\!\! \quad . \label{eq:defeigen}
\end{eqnarray}
Upon substituting Eq.(\ref{eq:defdelta}) in Eq.(\ref{eq:defgreen})
and using the eigenvalue Equation (\ref{eq:defeigen}), we can
write down the Green function as
\begin{eqnarray}
\Gamma({\bf u},{\bf u}') = - R^2 \sum _{l=1} ^{\infty} \sum _{m=-l}
^{m=+l} \!\!\!\! \quad \frac{Y^{m}_{l}(\theta,\phi) {Y^{m} _{l}}
^{\ast} (\theta',\phi')}{l(l+1)} \!\!\!\!\! \quad .
\label{eq:invert2}
\end{eqnarray}
We have used the fact that $Y^{0}_{0}=\sqrt{\frac{1}{4\pi}}$, and used the neutralizing background charge $\frac{1}{S}$ in
Eq.(\ref{eq:defgreen}) to cancel out the $l=0$ diverging mode. 

To simplify the series in Eq.(\ref{eq:invert2}), we exploit the
familiar identity \cite{Wyld-book}
\begin{eqnarray}
\sum _{m=-l} ^{m=+l} \!\!\!\! \quad Y^{m}_{l}(\theta,\phi) {Y^{m}
_{l}} ^{\ast} (\theta',\phi') = \frac{2l+1}{4 \pi} \!\!\!\!\!
\quad P_{l}(\cos \beta) \!\!\!\!\! \quad , \label{eq:identity1}
\end{eqnarray}
where $\beta$ is the angle (relative to the center of the sphere) between the directions
$\{\theta,\phi\}$ and $\{\theta',\phi'\}$ (see also
Eq.(\ref{eq:defectint2})). Upon substituting
Eq.(\ref{eq:identity1}) in Eq.(\ref{eq:invert2}), we find
\begin{eqnarray}
\Gamma({\bf u},{\bf u}') = - \sum _{m=-l} ^{m=+l} \!\!\!\! \quad
\left(\frac{1}{l}+\frac{1}{l+1}\right) \!\!\!\!\! \quad \frac{P_{l}(\cos
\beta)}{4 \pi} \!\!\!\!\! \quad . \label{eq:invert3}
\end{eqnarray}
The first term of the sum in Eq.(\ref{eq:invert2}) can be
simplified using the following identity \cite{Grad-book}
\begin{eqnarray}
\sum _{l=l} ^{l=\infty} \!\!\!\! \quad \frac{P_{l}(\cos \beta)}{l}
  &=&  -
\ln \left[1 + \sin \left(\frac{\beta}{2}\right) \right] \nonumber
\\ &-& \ln \left[\sin\left( \frac{\beta}{2}\right)\right]  \!\!\!\! \quad ,
\label{eq:identity2}
\end{eqnarray}
while for the second term we substitute
\begin{eqnarray}
\sum _{l=l} ^{l=\infty} \!\!\!\! \quad \frac{P_{l}(\cos \beta)}{l+1}
 &=&
\ln \left[1+ \left(\sin \frac{\beta}{2}\right)\right] \nonumber \\
&-& \ln \left[\sin \left(\frac{\beta}{2}\right) \right] -1
 \!\!\!\!\! \quad , \label{eq:identity2}
\end{eqnarray}
with the result
\begin{eqnarray}
\Gamma({\bf u},{\bf u}') = \frac{1}{4 \pi}
\ln\left[\frac{1-\cos\beta}{2} \right] + \frac{1}{4 \pi}
\!\!\!\!\! \quad . \label{eq:invert4}
\end{eqnarray}
Upon dropping additive constants that do not contribute to the
energy and substituting in Eq.(\ref{Coulomb model-energy}) we
obtain
\begin{equation}
F = -\frac{\pi K}{2 p^2} \sum_{i \neq j} q_{i} q_{j} \ln \left[1-
\cos \beta_{ij} \right] + \sum_{i=1}^{N_d} q_{i}^2 E_c \!\!\!\!
\quad , \label{eq:defectint1b}
\end{equation}
where the phenomenologically determined core energy $E_c$ has
been added by hand and reflect the microscopic physics not
captured by our long-wavelength theory.

\section{\label{appB} Liquid crystal textures and conformal mappings}

This Appendix collects a number of results from the theory of
complex variables relevant to the study of liquid crystal textures. 
The perspective adopted is to link the
liquid crystal elasticity to the intrinsic geometry of the texture
by the use of conformal transformations. The same method provides
an elegant route to finding the flow lines of simple
incompressible fluids in 2D \cite{feynmanbook} and to the exact
solution of analogous problems in electromagnetism and elasticity
\cite{Needham-book,Smirnov3.2}.

Nematic textures in the plane in the one Frank constant approximation
can be obtained by solving Laplace equation, which is conformally invariant, and in complex
coordinates $\{z=x+iy,\overline{z}=x-iy\}$ reads
\begin{eqnarray}
\partial _{\overline{z}} \!\!\!\! \quad \partial_{z} \!\!\!\!\! \quad \alpha = 0 \!\! \quad .
\label{eq:complex-laplacian}
\end{eqnarray}
Here $\alpha(x,y)\equiv \alpha(z,\overline{z})$ is the bond angle that the director
$\textbf{n}=(\cos{\alpha},\sin{\alpha})$ forms with respect to a fixed
direction, say the real axis $\widehat{x}$. The flat space laplacian equation (\ref{eq:complex-laplacian}) is also obtained by minimizing the
free energy of a vector field on a sphere (see Appendix
\ref{appA}) provided that a "stereographic projection gauge" is
chosen to carry out the calculation \cite{lube92,ovru91}. The
stereographic projection maps an arbitrary point on the sphere
$\textbf{R}(\theta,\phi)$ to the corresponding point $z=2R
\tan\left(\frac{\theta}{2} \right)e^{i\phi}$ in the complex plane.
The metric reads \cite{lube92}
\begin{equation}
g_{ij} = \frac{1}{2 \left(1 + \frac{z \bar{z}}{4R^2}  \right)} \begin{pmatrix} 0 & 1 \\
  1 & 0
\end{pmatrix} \!\!\!\! \quad ,
\label{eq:metric-mat}
\end{equation}
and the components of the gauge field in Eq.(\ref{eq:patic-ener}) are given by
\begin{equation}
A_{z} =\bar{A}_{\bar{z}}=-\frac{1}{2iz} \left(\frac{1- \frac{z \bar{z}}{4R^2}}{1 + \frac{z \bar{z}}{4R^2}}\right)
\!\!\!\! \quad .
\label{eq:A}
\end{equation}
In this representation of liquid crystal order on a sphere, the Frank free energy is
\begin{equation}
F = \frac{K}{4} \int d^2 z
\!\!\!\! \quad |\partial_{z}\alpha - A_{z}|^2 \!\!\!\! \quad , \label{eq:F}
\end{equation}
and the corresponding Euler Lagrange equation is indeed
Eq.(\ref{eq:complex-laplacian}), since the divergence of the gauge field is
zero ($\partial_{z}A_{\bar{z}}+\partial_{\bar{z}}A_{z}=0$) \cite{lube92}. The stereographic projection provides an
example of how conformal transformations can be used to map physics on an
arbitrary curved surface onto simpler planar problems. This technique
can also be employed to analyze two dimensional order on surfaces
of $varying$ Gaussian curvature (see Ref.\cite{ViteNels04,ViteTurn05}).

\begin{figure}
\includegraphics[width=0.45\textwidth]{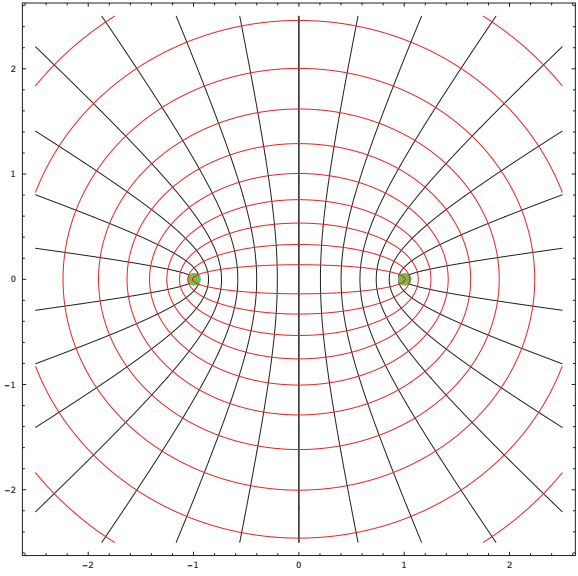}
\caption{(Color online) Splay rich (hyperbolic) and bend-rich (ellipsoidal) families of nematic "flow" lines generated
by two $s=\frac{1}{2}$ disclinations. The two families of flow lines
are the equipotential and field lines of a complex function
$\Omega(z)$ whose branch cut is a horizontal line connecting the two
defects.\label{fig:conformal-plot-app}}
\end{figure}

A second use of conformal mappings is as generators of 2D liquid
crystal textures in bounded geometries or in the presence of
defects. Consider an analytic function $w=f(z)$ that maps a grid of horizontal and vertical
lines in the complex plane $z=x+iy$ onto a family of orthogonal curves
in the $w=u+iv$ plane that are respectively streamlines and
equipotential lines of the corresponding flow (see Fig.
\ref{fig:conformal-plot-app}).
Similarly, we can define the inverse function $\Omega(z)=f^{-1}(z)$
and note that $\Omega(z)=\Phi(x,y)+i\Psi(x,y)$ maps equipotential lines and streamlines, given by the contour
lines of $\Phi(x,y)$ and $\Psi(x,y)$, into a grid of vertical and
horizontal lines respectively.

The connection between liquid crystal textures and conformal mappings rests on the following observation:
if the director $\textbf{n}(z)$ forms a constant angle with respect to the streamlines (or the equipotential lines) of $\Omega(z)$,
then $\alpha(z)$ automatically
satisfies Eq.(\ref{eq:complex-laplacian}) \cite{Needham-book}.
In the one Frank
constant approximation, the complex nematic director $\vec{n}(z)=n_{x}(x,y)+i n_{y}(x,y)$ is given (up to an
arbitrary global rotation) by
\begin{eqnarray}
\vec{n}(z) = \frac{\overline{\Omega '(z)}}{|\overline{\Omega
'(z)}|} \!\! \quad , \label{eq:director}
\end{eqnarray}
The complex function $\Omega '(z)$ denotes the
derivative with respect to $z$ of the complex function $\Omega=\Phi(x,y)+i\Psi(x,y)$
and we define $\overline{\Omega} \equiv \Phi(x,y)-i\Psi(x,y)$.
The bond angle is readily expressed in terms of
$\Omega(z)$ via the relation
\begin{eqnarray}
\alpha(z) = - \Im m \log\left[\Omega '(z)\right] \!\! \quad ,
\label{eq:bond1}
\end{eqnarray}
where $\Im m$ denotes the imaginary part of a complex number.

As an illustration consider the simple case of two disclinations
on the real axis at positions $x=\pm1$ respectively. The nematic
director rotates by $\pi$ on a path encircling only one defect and
by $2\pi$ on a path enclosing both (see Fig.
\ref{fig:conformal-plot-app}). These requirements are met by
choosing the complex function $\Omega(z)=\arccos(z)$ that is
analytic everywhere except for a branch cut on the real axis from
$x=-1$ to $x=1$. The streamlines and equipotential lines are a
family of hyperbolas and ellipses with coinciding foci at $x=\pm
1$; they correspond to two distinct nematic textures dominated
by splay and bend respectively.
%%%%%%%%%%%%%%%%%%%%%%%%%%%%%%%%%%
%%%%%%%%%%%%%%%%%%%%%%%%%%%%%%%%%%%
The bond angle of the director oriented along the streamlines is
easily extracted from the argument of the complex vector field in
Eq.(\ref{eq:director}), with the result
\begin{equation}
\alpha(x,y) =\arctan \left[ \frac{y^2 + 1 - x^2 - \sqrt{(y^2 + x^2
+ 1)^2 -4x^2}}{2 x y}\right] \!\!\!\!\!\! \quad .
\label{eq:thetaecplicit}
\end{equation}
In this simple example, one can explicitly check that the result
in Eq.(\ref{eq:thetaecplicit}) is recovered through the more
familiar route of superposing the solutions corresponding to the
two isolated defects
\begin{equation}
\alpha(x,y) =\frac{1}{2}\arctan \left[ \frac{y}{x-1}\right]+
\frac{1}{2}\arctan \left[ \frac{y}{x+1}\right] \!\!\!\!\!\! \quad
. \label{eq:thetasecondway}
\end{equation}

The applicability of the method of conformal mappings to finding
liquid crystal textures can be justified by means of simple
geometric reasoning. We start by noting that the curl and
divergence of a two dimensional vector field $\bf{v}$, whose streamlines and
orthogonal trajectories are labelled by the subscripts $s$ and $p$
respectively, can be expressed geometrically via the relations
\cite{Needham-book}
\begin{equation}
\partial _x \!\!\!\!\! \quad
v_{x}+ \partial _y \!\!\!\!\! \quad
v_{y} \equiv \textbf{$\nabla$} \cdot \textbf{v} = \partial _s \!\!\!\!\! \quad
|v| + \kappa _p \!\!\!\!\! \quad |v| \!\! \quad . \label{eq:div}
\end{equation}
\begin{equation}
\partial _x \!\!\!\!\! \quad
v_{y} - \partial _y \!\!\!\!\! \quad
v_{x} \equiv \textbf{$\nabla$} \times \textbf{v} = - \partial _p \!\!\!\!\!
\quad |v| + \kappa _s \!\!\!\!\! \quad |v| \!\! \quad ,
\label{eq:curl}
\end{equation}
where $\kappa _s$ and $\kappa _p$ are the respective curvatures
while $\partial_s$ and $\partial_p$ are the directional
derivatives along the two orthogonal families of level curves.
The direction of increasing $p$ is chosen to make a
counterclockwise $\frac{\pi}{2}$ angle with $\bf{v}$. For
example the black lines in Fig. \ref{fig:conformal-plot-app} trace
the electric field $\bf{v}$ generated by a uniformly charged plate
or the flow lines of an ideal fluid exiting a slit of width
given by the branch cut. Unlike the liquid crystal director in
Eq.(\ref{eq:director}), the magnitude of $\bf{v}$ is allowed to
vary with position. By construction, such a vector field is
divergence free and curl free, hence
\begin{equation}
\kappa _s = \partial _p \log|v| \!\! \quad , \label{eq:kappa-s}
\end{equation}
\begin{equation}
\kappa _p = - \!\!\!\! \quad \partial _s \log|v| \!\! \quad .
\label{eq:kappa-p}
\end{equation}
By combining Equations (\ref{eq:kappa-s}) and (\ref{eq:kappa-p}),
we obtain the geometric condition that a family of equipotential
lines (or streamlines) needs to satisfy in order to be identified
as level curves of an harmonic potential, namely
\begin{equation}
\frac{\partial \kappa _p}{\partial p} + \frac{\partial \kappa
_s}{\partial s}=0 \!\! \quad . \label{eq:rate}
\end{equation}
This condition is entirely cast in terms of the curvatures of the
equipotential lines and streamlines without explicit reference
to either the potential to be assigned or to the magnitude of
the vector field $\textbf{v}(z)$ \cite{Bive92,Need94}. This is a
natural language to discuss orientational order in liquid crystals since the director $\vec{n}(z)$ is a vector field of unit
magnitude.

If we take the liquid crystal director to form a constant angle
with respect to $\textbf{v}(z)$, the curvatures $\kappa_ s$ and
$\kappa_ p$ in Eq.(\ref{eq:rate}) can be simply cast as the
directional derivatives of $\alpha$ along the streamlines and
equipotential lines respectively. In fact, the curvature of these
contour lines is the rate of change of their directions which is
naturally parameterized by $\alpha$. Hence Eq.(\ref{eq:rate})
reduces to
\begin{equation}
\frac{\partial ^2 \alpha}{\partial p ^2 } + \frac{\partial ^2
\alpha}{\partial ^2 s}=0 \!\! \quad . \label{eq:laplacian21}
\end{equation}
The left hand side of Eq.(\ref{eq:laplacian2}) is the Laplacian of
$\alpha$ expressed in terms of orthogonal coordinates along $s$
and $p$. Since the Laplacian is coordinate independent,
Eq.(\ref{eq:laplacian21}) is equivalent to
Eq.(\ref{eq:complex-laplacian}) and $\alpha (x,y)$ represents
(apart from an arbitrary global rotation) the desired texture that
minimizes the Frank free energy when $K_1=K_3$.

As a byproduct, Equations (\ref{eq:kappa-s}) and
(\ref{eq:kappa-p}) can help to visualize how the elastic energy
stored in every portion of the texture of Fig.
\ref{fig:conformal-plot-app} is distributed between bend and splay
deformations. For most liquid crystals $K_3 > K_1$, so the texture
with director tangent to the streamlines will be energetically
favored (black lines in Fig. \ref{fig:conformal-plot-app}). In
this case, the full Frank energy density can be rewritten in terms
of the local curvatures of streamlines and equipotential lines via
the simple relation
\begin{equation} K_3 \!\!\!\!
\quad(\textbf{$\nabla$} \times \textbf{n})^2 + K_1 \!\!\!\! \quad
(\textbf{$\nabla$} \cdot \textbf{n})^2 = K_3 \!\!\!\! \quad
{\kappa_s}^2 + K_1 \!\!\!\! \quad {\kappa_p}^2 \!\! \quad .
\label{eq:fullfrank}
\end{equation}
The energetically costly deformation involving bend takes place
only around the defects in a region of radius of the order of
their separation. Elsewhere $\kappa_s$ is vanishingly small.
In contrast, $\kappa_p$ drops off more slowly at large distances like the inverse
of the radius of a circle centered on the midpoint between the two
disclinations. Splay deformations are present throughout the
system but they have a smaller energy cost $K_1$. The converse
situation occurs if $K_3 < K_1$ so that the texture represented by
the red equipotential lines in Fig. \ref{fig:conformal-plot-app}
becomes energetically favored.

The curvatures $\kappa_s$ and $\kappa_p$ are respectively the real
and imaginary parts of the complex curvature, $K(z)=\kappa_{s}+i\kappa_{p}$, of the
mapping. This quantity can be readily derived from the complex potential
$\Omega(z)$. For calculational purposes, it is more
convenient to recast Eq.(\ref{eq:kappaz}) in the form $\kappa_p + i \kappa_s = -\frac{|\Omega'| \!\!\!\!\! \quad \Omega''}{\Omega'^2}$.
\begin{equation}
K(z)\equiv\kappa_s + i \kappa_p = - i \!\!\!\!\! \quad
\frac{|\Omega'| \!\!\!\!\! \quad
\overline{\Omega''}}{\overline{\Omega'} \!\!\!\!\! \quad ^2} \!\!
\quad . \label{eq:kappaz}
\end{equation}
The reader is referred to the mathematical literature
\cite{Bive92,Needham-book} for a proof of Eq.(\ref{eq:kappaz}).
The intuitive significance of the complex curvature can be grasped
by considering how a conformal mapping $f=\Omega ^{-1}$ acts on a
curve with local curvature $\kappa$ at a point in the $z=x+iy$
plane. The curve is mapped onto an image curve in the $w=u+iv$
plane whose curvature $\kappa'$ at the corresponding point differs
from $\kappa$. The curvature $\kappa'$ of the image curve is
determined by two mechanisms. Firstly, the mapping $f(z)$ locally
stretches distances by a factor $|f'(z)|$, hence the radius of
curvature of the image curve will be naturally multiplied by this
amplifying factor. The second mechanism arises because a
conformal mapping can introduce curvature even if none was
originally present (in the isothermal net) simply by locally
twisting the direction of the isothermal net by an angle equal to
$\arg \!\!\!\! \quad [f'(z)]$. For this reason, the
complex derivative $f'(z)$ is sometimes called an amplitwist and
encodes information on the local effect of the mapping \cite{Needham-book}. The
non-analytic function $K(z)$ controls the amount of curvature
generated ex-novo by the mapping. For example, the mapping
$f(z)=Cos(z)$ transforms a grid of horizontal lines (think of them
as a possible direction for the nematic molecules in the
defect-free ground state) into the family of hyperbolae in Fig.
\ref{fig:conformal-plot-app} corresponding to a defected texture
with two $+1/2$ disclinations. It is not surprising that the free
energy density stored in the defected texture is simply
proportional (in the one Frank constant approximation) to
\begin{equation}
\left(\frac{\partial \alpha}{\partial p}\right)^2 +
\left(\frac{\partial \alpha}{\partial s}\right)^2 = \kappa_s ^2 + \kappa_p ^2 = |K(z)|^2 \!\! \quad , \label{eq:laplacian2}
\end{equation}
where the two elastic constants were set to be equal in Eq.(\ref{eq:fullfrank}) and
the director $\textbf{n}(z)$ was parameterized
in terms of the bond angle $\alpha(z)$.
The Frank free energy is thus proportional to the complex curvature modulus-squared
in analogy with the Helfrich
free energy of a membrane whose derivation rests on an higher
dimensional generalization of Eq.(\ref{eq:fullfrank}).

\section{\label{appC} Vibrational spectrum of colloidal molecules}

\begin{figure}
\includegraphics[width=0.45\textwidth]{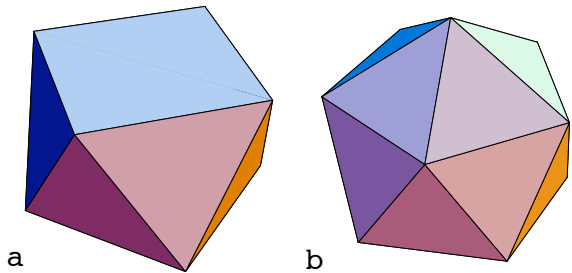}
\caption{\label{fig:modes}(a) The ground state of a tetratic phase exhibits eight short disclination lines located at the vertices of a square antiprism inscribed in the sphere. 
(b) The twelve disclination that characterize hexatic order s sphere 
lie at the vertices of an icosahedron.}
\end{figure}

In this appendix we provide an introduction to the group theoretical treatment of the
vibrational spectrum of colloidal "molecules". The more complicated cases of hexatic (p=6) and tetratic (p=4) order 
are analyzed in some detail (see Fig. \ref{fig:modes}). 
%%%%%%%%%%%%%%%%%%%%%%%%%%%%%%%
%%%%%%%%%%%%%%%%%%%%%%%%%%%%%%%

The starting point of the group theoretical treatment of the
vibrational spectrum of colloidal "molecules" is the observation
that the defects displacements from equilibrium, $\textbf{q}$ (see
Eq.(\ref{eq:coords}), form the basis of a $reducible$
representation of the point group of the molecule. 
If a molecule
is acted upon by a symmetry operation, a new configuration will
result in which the displacements of each defect will be permuted
and transformed, but inter-defect distances and angles
will be preserved. Here we take the point of view that the
defects themselves are not permuted only their displacements, for
example defect $i$ may exchange its displacement coordinates
$q_{i}$ with defect $j$. The liquid crystal free energy (in the one
Frank constant approximation) is therefore invariant under all
the operations of the point group of the colloidal defect array.

The action of each operation of the group is naturally
represented by a distinct $2N \times 2N$ matrix ($N$ is the
number of defects) that relates the new and old defect positions.
This representation can be completely reduced by choosing a set 
of symmetry-related normal coordinates 
that are obtained from the original ones by
means of a linear transformation. When normal coordinates are used,
the matrixes representing the action of the symmetry group can be brought
in block diagonal form simultaneously. Energetically degenerate linear combination
of the original coordinates form the smallest sets invariant
under application of any symmetry operation of the
group. The members of any one set generate an
$irreducible$ representation of the group.
%%%%%%%%%%%%%%%%%%%%%%%%%%%%%%%%%
\begin{table}
\caption{\label{tab:icosahedron} Character for the irreducible
representations of the icosahedral point group together with the
character of the twenty-four-dimensional representation $\Upsilon$
generated by the defect displacements.}
\begin{ruledtabular}
\begin{tabular}{lccccr}
$Y$ &  $E$ & 12$C_5$ & 12$C^{2}_{5}$ & 20$C_3$ & 15$C_{2}$ \\
\hline
$A$ & 1 & 1 & 1 & 1 & 1 \\
$F_1$ & 3 & $\tau$ \footnote{Note that $\tau=\frac{\sqrt{5}+1}{2}$.} & $- \!\!\!\! \quad \tau^{-1}$ & 0 & $-1$ \\
$F_2$ & 3 & $- \!\!\!\! \quad \tau^{-1}$ & $\tau$ & 0 & $-1$ \\
$G$ & 4 & $-1$ & $-1$ & 1 & 0 \\
$H$ & 5 & 0 & 0 & $-1$ & 1 \\
 &  &  &  &  &  \\
$Y$ & 24 & $2 \tau ^{-1}$ & $-2 \tau$ & 0 & 0 \\
\end{tabular}
\end{ruledtabular}
\end{table}
%%%%%%%%%%%%%%%%%%%%%%%%%%%%%%%%%
For each point group there is only a small number of
inequivalent irreducible representations generally classified by the
characters of their transformations. The characters of the
transformations are simply defined as the traces of the matrices
corresponding to each symmetry operation and they are conveniently
tabulated in most texts of group theory
\cite{WilsonDeciusCross-book,Landau-quantum-book} (see Tables
\ref{tab:tetrahedron}-\ref{tab:tiltedcube} for the character
tables relevant to the tetrahedral, icosahedral and twisted-cube
shaped distributions of defects on a sphere).

The task of finding the number of degenerate eigenmodes, $n^{(\gamma)}$, with a
given symmetry (labelled by $\gamma$) reduces to counting
how many times the corresponding irreducible representation
appears in a reducible
representation. Note that
the characters of the original representation are the same as the
ones of the completely reduced one since the two differ only by a
change of coordinates which preserves the trace. Thus, the character $\chi _{R}$ of the completely reduced
representation will be the sum of the characters of the various
irreducible representations that it contains
\begin{equation}
\chi _{R} = \sum _{\gamma} n^{(\gamma)} \chi_{R} ^{(\gamma)} \!\!
\quad , \label{eq:rep}
\end{equation}
where $\chi_{R} ^{(\gamma)}$ labels the character of the symmetry
operation $R$ in the irreducible representation $\gamma$. By
appealing to the orthogonality of the characters one can write an
expression for $n^{(\gamma)}$ in analogy with the familiar
expression for the component of a vector along a given basis axis \cite{WilsonDeciusCross-book,Landau-quantum-book}
\begin{equation}
n^{(\gamma)} = \frac{1}{g} \sum _{R}  \chi _{R} ^{(\gamma)\ast}
\!\!\!\!\! \quad \chi_{R} \!\! \quad , \label{eq:rep2}
\end{equation}
where $g$ is the number of the symmetry operations in the group
and $\chi_{R}$ is the character of the completely reduced
representation. Eq.(\ref{eq:rep2}) is equivalent to
Eq.(\ref{eq:group1}) quoted in the main text, as long as the sum
over the group elements $R$ is replaced by a weighted sum
over the classes in the group since the characters of group
elements in the same class are equal.

We now adopt the analysis of Ref.\cite{WilsonDeciusCross-book,Landau-quantum-book} 
to provide a set of rules that produce the
characters $\chi_{R}$ of the reducible representation generated 
by the coordinates
without working out the full form of the transformation matrices.
There are two key points to
notice. First only the defects located on a symmetry axis or
plane contribute to $\chi_{R}$ the trace of the transformation matrix; 
defects whose displacements are instead interchanged or permuted by the symmetry
operation contribute only to the non-diagonal terms of the matrix
and hence can be ignored in determining the character $\chi_{R}$. Second
the directions along which the displacements from equilibrium are
measured can be chosen freely since the trace is invariant upon
coordinate transformations. It is generally convenient to
choose them so that only one of the two displacements components
is affected by the symmetry operation.
%%%%%%%%%%%%%%%%%%%%%%%%%%%%%%%%%%%%%
\begin{figure}
\includegraphics[width=0.45\textwidth]{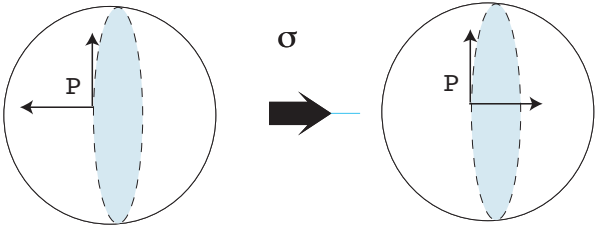}
\caption{\label{fig:reflection} Schematic illustration of the
inversion through an equatorial plane of symmetry (shaded circle bisecting a sphere) for a reflection $\sigma$. The
longitudinal displacement of the defect is unchanged while the
direction of the latitudinal one is inverted.}
\end{figure}
%%%%%%%%%%%%%%%%%%%%%%%%%%%%%%%%%%%%%%

As a simple example, consider a defect lying on a reflection plane
(see Fig. \ref{fig:reflection}). The displacement vectors before
and after the symmetry operation is applied are $\{\delta
\theta,\delta \phi \}$ and $\{\delta \theta , -  \delta \phi \}$
respectively. The resulting contribution to the character from a
single defect is thus $1-1=0$. Inversions through a center of symmetry 
(coinciding with the center of the sphere)
have a vanishing contribution to $\chi_{R}$ because there are no 
defects there that can contribute to $\chi_{R}$.

A rotation $C_{n} ^{k}$ by an angle $\frac{2 \pi k}{n}$ through an n-fold
axis of symmetry on which the defect lies leads to the following
transformation laws for the longitudinal and latitudinal
displacements
\begin{equation}
\left(%
\begin{array}{c}
\delta \theta _{i} \!\!\!\!\! \quad ' \\
\\
\delta \phi _{i} \!\!\!\!\! \quad '  \\
\end{array}%
\right) = \left(%
\begin{array}{cc}
\cos \left(\frac{2 \pi k}{n} \right)    & - \sin \left(\frac{2 \pi k}{n} \right) \\
\\
  \sin \left(\frac{2 \pi k}{n} \right) & \cos \left(\frac{2 \pi k}{n} \right) \\
\end{array}%
\right)  \left(%
\begin{array}{c}
\delta \theta _{i}  \\
\\ \delta \phi _{i}   \\
\end{array}%
\right) \!\! \quad .
\label{eq:rot}
\end{equation}
We measure displacements using polar coordinates with respect to the symmetry axis; the prime denotes the orthogonal displacements after the
symmetry operation $C_{n} ^{k}$ is applied. Inspection of
Eq.(\ref{eq:rot}) shows that the contribution from $C_{n} ^{k}$ to
the character $\chi_{R}$ is equal to $2 \cos \left(\frac{2 \pi
k}{n}\right)$ times the number of defects lying on the axis of
rotation. On the other hand the contribution to $\chi_{R}$  from
the improper rotation $S^{k} _{n}$ is zero. To see this note that the symmetry operation
$S^{k} _{n}$ is a rotary reflection achieved by performing a
successive rotation through an (alternating) axis followed by a
reflection in the plane perpendicular to the axis $k$ times. An example of a
molecule possessing the symmetry operation $S_{4}$ is methane,
CH$_{4}$, with the carbon atom lying at the intersection between
an alternating axis and the reflection plane (See Fig.
5-2 in Ref. \cite{WilsonDeciusCross-book}). The
tetrahedral defect configuration considered in this paper does not
posses any defect at the position occupied by the carbon atom of
the methane molecule. More generally, the possibility of having a
defect whose equilibrium position is unchanged by the rotary
reflection is ruled out because such defect would have to lie off
the spherical surface at the intersection between the alternating
axis and the plane of reflection.

To sum up, each of the characters, $\chi_{R}$, of the completely
reduced representation formed by the displacement coordinates is
given by the number of atoms whose equilibrium positions are not
changed by the symmetry operation $R$ times its fundamental
character as derived in the previous paragraphs. Similar
results that apply to unconstrained molecules whose atoms have
three dimensional displacements are listed in Table 6-1 of Ref.
\cite{WilsonDeciusCross-book}. The resulting characters for the
tetrahedral, icosahedral and tilted cube defects configurations
are listed in Tables \ref{tab:tetrahedron}-\ref{tab:tiltedcube}.
%%%%%%%%%%%%%%%%%%%%%%%%%%%%%%%%
\begin{table}
\caption{\label{tab:tiltedcube} Character for the irreducible
representations of the tilted cube point group together with the
character of the sixteen-dimensional representation $\Xi$
generated by the defect displacements.}
\begin{ruledtabular}
\begin{tabular}{lccccccr}
$D_{4d}$ &  $E$ & 2$S_8$ & 2$C_{4}$ & 2$S^{3}_{8}$ & $C_{2}$ & 4$C^{1}_{2}$ & 4$\sigma_{2}$ \\
\hline
$A_1$ & 1 & 1 & 1 & 1  & 1 & 1 & 1 \\
$A_2$ & 1 & 1 & 1 & 1  & 1 & $-1$ & $-1$ \\
$B_1$ & 1 & $-1$ & 1 & $-1$  & 1 & $1$ & $-1$ \\
$B_2$ & 1 & $-1$ & 1 & $-1$  & 1 & $-1$ & $1$ \\
$E_1$ & 2 & $\sqrt{2}$ & 0 & $-\sqrt{2}$  & $-2$ & $0$ & $0$ \\
$E_2$ & 2 & $0$ & $-2$ & $0$  & 2 & $0$ & $0$ \\
$E_3$ & 2 & $-\sqrt{2}$ & $0$ & $\sqrt{2}$  & $-2$ & $0$ & $0$ \\
&  &  &  &  &  &  & \\
$\Xi$ & 16 & 0 & 0 & 0 & 0 & 0 & 0 \\
\end{tabular}
\end{ruledtabular}
\end{table}
%%%%%%%%%%%%%%%%%%%%%%%%%%%%%%%
Upon using Eq.(\ref{eq:group1}) and the character table
\ref{tab:icosahedron} we can decompose the 24 dimensional
representation, $Y$, formed by the displacements from an
icosahedral equilibrium configuration into irreducible
representations. The result reads
\begin{equation}
Y = 2 H + 2 G + 2 F_{1} \!\! \quad .
\label{eq:decomposeico}
\end{equation}
The three rigid body rotations correspond to
one of the two triplets in $F_{1}$ while the remaining 21
independent normal coordinates form energetically degenerate
multiplets with the following degeneracy factors: 2 quintets, 2
quartets and 1 triplet. This analysis is confirmed upon direct
diagonalization of the representation $Y$ which leads the
21 non-vanishing eigenvalues $\lambda_{i}$, with the multiplicities
shown bold in parenthesis
\begin{eqnarray}
\lambda_{i} =\{0.87 \!\!\!\!\!\!\! \quad \times (\textbf{5}),
\!\!\!\!\! \quad 0.09 \!\!\!\!\!\!\! \quad \times (\textbf{5}),
\nonumber\\ 
\!\!\!\!\! \quad 0.74 \!\!\!\!\!\!\! \quad \times (\textbf{4}),
\!\!\!\!\! \quad 0.22 \!\!\!\!\!\!\! \quad \times (\textbf{4}),
\!\!\!\!\! \quad 0.96 \!\!\!\!\!\!\! \quad \times (\textbf{3})\}
 \!\! \quad . \label{eq:decomposeico}
\end{eqnarray}
Note that the normal
modes in the second quintet are much "softer" than the rest.

A similar analysis applied to the twisted cube configuration of
defects leads to the following decomposition of the (defects'
displacements) representation, $\Xi$,
\begin{equation}
\Xi = 2 E_{1} + 2 E_{2} + 2 E_{3} + A_{1} + A_{2} + B_{1} + B_{2}
\!\! \quad . \label{eq:decomposetilt}
\end{equation}
where the three rigid body rotations are contained in one of the
two doublets $E_{3}$ and in the singlet $A_{2}$, which leaves five
doublets and three singlets for the eigenvalues $\zeta_{i}$.
Direct diagonalization of $\Xi$ leads four doublets, one triplet
and two singlets of non-vanishing eigenvalues,
\begin{eqnarray}
\zeta_{i} =\{1.37 \!\!\!\!\!\!\! \quad \times (\textbf{3}),
\!\!\!\!\! \quad 1.31 \!\!\!\!\!\!\! \quad \times (\textbf{2}),
\!\!\!\!\! \quad 0.89 \!\!\!\!\!\!\! \quad \times (\textbf{2}),
\nonumber\\ 
\!\!\!\!\! \quad 0.47 \!\!\!\!\!\!\! \quad \times (\textbf{2}),
\!\!\!\!\! \quad 0.06 \!\!\!\!\!\!\! \quad \times (\textbf{2}),
\!\!\!\!\! \quad 0.11 , \!\!\!\!\! \quad 1.26 \}
  \label{eq:decomposetilt}
\end{eqnarray}
The discrepancy between the degeneracies found by direct
diagonalization on one hand and group theory on the other is
caused by an accidental symmetry of the potential energy of the
tilted-cube arrangement of defects. Hence the first triplet is to
be interpreted as the missing doublet and singlet that happen to
have the same energy even if there is no symmetry reasons to
expect so. The modes in the last doublet of
Eq.(\ref{eq:decomposetilt}) are the softest.

\end{document}